\documentstyle[pre,aps,epsfig]{revtex}
\begin{document}

\draft

% for two column  activate the line below...
%\twocolumn[\hsize\textwidth\columnwidth\hsize\csname@twocolumnfalse\endcsname
\title{
Interruption of torus doubling bifurcation and genesis of strange nonchaotic
attractors in a quasiperiodically forced map : Mechanisms and their 
characterizations}

\author{A. Venkatesan and M. Lakshmanan}
\address{Centre for Nonlinear dynamics,\\ Department of Physics,\\
Bharathidasan University,\\Tiruchirapalli 620 024, INDIA }

\maketitle

\begin{abstract}

A simple quasiperiodically forced one-dimensional cubic map is shown to
exhibit very many types of routes to chaos via strange nonchaotic
attractors (SNAs) with reference to a two-parameter $(A-f)$ space.  The
routes include transitions to chaos via SNAs from both one frequency
torus and period doubled torus. In the former case, we identify the
fractalization and type I intermittency routes. In the latter case, we
point out that atleast four distinct routes through which the
truncation of torus doubling bifurcation and the birth of SNAs take
place in this model. In particular, the formation of SNAs through
Heagy-Hammel, fractalization and type--III intermittent mechanisms are
described. In addition, it has  been found that in this system there
are some regions in the parameter space where a novel dynamics
involving a sudden expansion of the attractor which tames the growth of
period-doubling bifurcation takes place, giving birth to SNA.  The SNAs
created through different mechanisms are characterized by the behaviour
of the Lyapunov exponents and their variance, by the estimation of
phase sensitivity exponent as well as through the distribution of
finite-time Lyapunov exponents.

\end{abstract}
%\tightenlines
%\narrowtext
%\vskip1pc]
\section{ INTRODUCTION}

Torus doubling bifurcation (geometrically similar to period doubling
bifurcation) as a universal route to chaos is turning out to be one of
the leading topics of research in the  study of quasiperiodically
forced chaotic dynamical systems  during the past few years \cite
{R1,R2,R3,R4,R5,R6}. The existence of such an exotic bifurcation in
several experimental situations and theoretical models indicates the
importance of this bifurcation in improving our understanding of the
qualitative and quantitative behaviours of dynamical systems \cite
{R1,R2,R3,R4,R5,R6,R7,R8,R9,R9a,R10}.  A very common observation is
that such systems do not undergo an infinite sequence of doubling
bifurcations as in the case of the lower dimensional systems; instead,
the truncation of torus doubling begins when the doubled torus becomes
extremely wrinkled and then gets destroyed.  Such a destroyed torus is
a geometrically strange (fractal dimensional) object in the phase
space, a property that usually corresponds to that of a chaotic
attractor. However, it would not exhibit sensitivity to initial
conditions asymptotically (for example, Lyapunov exponents are
nonpositive) and hence is not chaotic and so it is a strange nonchaotic
attractor(SNA)
\cite{R11,R12,R12a,R12b,R12c,R12d,R13,R13a,R13b,R14,R16,R16a,R17,R18,R19,R20,R21,R22}.
Actually the existence of SNA was first identified by Grebogi {\it et
al.} \cite {R11} in their work on the transition from two-frequency
torus to chaos via SNA.  Later on, it was found that these attractors
can arise in physically relevant situations such as a quasiperiodically
forced pendulum \cite{R12a,R16}, quantum particles in quasiperiodic
potentials \cite{R12a}, biological oscillators \cite{R12b},
Duffing-type oscillators \cite{R12c,R12d,R13}, velocity-dependent
potential systems \cite{R10}, electronic circuits \cite{R13a,R13b}, and
in certain maps \cite {R14,R16,R16a,R17,R18,R19} in different
transitions to SNA including the torus doubling bifurcation and the
birth of SNAs.  Also, the existence of torus doubling  truncation and
the appearance of SNA was confirmed by an experiment consisting of a
quasiperiodically forced, buckled  magnetoelastic ribbon \cite{R20}.
Besides this experiment, the exotic strange nonchaotic attractors were
studied in analog simulations of a multistable potential \cite{R21},
and in neon glow discharge experiment \cite {R22} through different
transitions to SNAs. The existence of SNAs in such physically relevant
systems has naturally motivated further intense investigations on the
nature and occurrence of them.

A question of intense further interest  in the topic is the way in
which the truncation of period doubling occurs to give birth to SNAs.
In particular, it has been found  that the birth of SNAs often occurs
due to the   collision of a period doubled torus with its unstable
parent so that a period 2$^k-$torus gives rise to a 2$^{k-1}-$band SNA
\cite {R8} or a gradual fractalization of torus, in which a period
2$^k-$torus approaches a 2$^k-$band SNA \cite {R9}. Recently, the
present authors have identified that the torus doubling sequence is
tamed due to a subharmonic bifurcation (subcritical period-doubling
bifurcation) leading to the creation of SNAs.  In addition, this
transition has been shown to exhibit type-III intermittent
characteristic scaling \cite {R12d,R13a}.  Apart from the creation of
SNAs due to the collapse of the tori, the authors have also shown that
there are some regions of the system parameters where the torus
doubling sequence is truncated by a merging bifurcation leading to the
formation of a torus bubble \cite {R10}, reminiscent of period bubbles
in low dimensional systems. Also, using the renormalization group
approach, Kuznetsov, Feudal and Pikovsky have revealed scaling properties
both for the critical attractor and for the parameter plane topography
near the terminal point of the torus doubling bifurcation \cite{R9a} in
connection with the above mentioned  collision scenario.

Besides the creation of SNAs through the above mentioned truncation of
torus doubling bifurcation, several other mechanisms have also been
studied in the literature for the birth of SNAs. The most common one is
gradual fractalization of a torus where an amplitude or phase
instability causes the collapse of the torus \cite{R9}. This is in fact
one of the least understood mechanisms for the formation of SNAs since
there is no apparent bifurcation unlike the torus collision mechanism
identified by Feudel, Kurths and Pikovsky where a stable torus and an
unstable torus collide at a dense set of points, leading to the
creation of SNAs \cite{R16a}.  Prasad, Mehra and Ramaswamy \cite{R14}
have shown that a quasiperiodic analogue of a saddle-node bifurcation
gives rise to SNAs through the intermittent route with the dynamics
exhibiting scaling behaviour characteristic of type--I intermittency.
Yalcinkaya and Lai have described  that an on-off intermittency can be
associated with SNA creation through a blow-out bifurcation when a
torus loses its transverse stability \cite{R13}. Other than the above
said scenarios, a number of other quasiperiodic routes to SNAs have
been described in the literature \cite{R12,R18}. They include the
existence of SNAs in the transition from two-frequency to
three-frequency quasiperiodicity \cite{R12}, transition from
three-frequency to chaos via SNA and transition  to chaos via strange
nonchaotic trajectories on torus \cite{R18}.

Considering particularly the different routes discussed above for the
inhibition of torus doubling sequence and the birth of SNAs, we note
that they have so far been identified essentially in $\it{different}$
dynamical systems. However,  it is important to study the truncation of
the torus doubling bifurcations and the appearance of SNAs in a single
system in order to  understand the mechanisms and their characteristic
features clearly. In this connection, we consider a simple model in the
form of an one-dimensional cubic map,
 $$x_{i+1}= - A x_i+x_i^3, \eqno(1) $$
which is quite analogous to the typical Duffing oscillator
\cite{R23,R24}. The existence of different dynamical features of this
system has been studied in refs.\cite{R24,R25}. In the present work, we
investigate the dynamics of (1) with the addition of a constant bias,
$$x_{i+1}= Q - Ax_i+x_i^3, \eqno(2) $$ 
and also subject to an additional quasiperiodic forcing, 
$$x_{i+1}= Q +f \cos (2 \pi \theta_i) - A x_i+x_i^3, \eqno(3) $$
$$\theta_{i+1}=\theta_i+ \omega (\text{mod} 1),$$ 
and show that the
latter  is a rich dynamical system in comparison with the former, 
possessing a vast number of
regular, strange nonchaotic and chaotic attractors in a two-parameter
$(A-f)$ space for a fixed $Q$. In particular, we focus our attention 
mainly on the
truncation of torus-doubling bifurcations leading to the birth of SNAs and the
mechanisms by which they arise in a range of the two-parameter $(A-f)$
space, besides pointing out the standard transitions to chaos via SNAs
from one  frequency torus. A variety of transitions from truncated
doubled torus to SNAs could  be identified, characterized and
distinguished in this system.

To start with, we show that the system (2) undergoes one or more period
doublings but it  need not complete the entire Feigenbaum cascade, and
that it may be possible to have only a finite number of period
doublings, followed by, for example, undoubling or other bifurcations
in the presence of constant bias,  as was shown by Bier and Bountis in
different systems \cite {R25}. The possibility of such a different
remerging bifurcation phenomenon in the torus doubling sequence is
reported in the present case, when the system (2)  is subjected to
quasiperiodic forcing as in model (3).  As the system (3) possesses
more than one control parameter and remains invariant under reflection
symmetry, the remergence  is likely to occur as in the absence of
quasiperiodic forcing in the system (2).  Our numerical study shows
that for fixed value of $Q$ in some regions of the $(A-f)$ parameter
space a torus doubled orbit emerges and remerges from a single torus
orbit at two different parameter values of $f$ to form a single torus
bubble. Such a remerging bifurcation can retard the growth  of the
torus doubled bifurcations and the development of the associated universal
route to chaos further. However, the nature of remerging torus doubled
bifurcation or,  more specifically, the torus bubbling ensures the
existence of different routes for the creation of SNAs when the full
range of parameters is explored. To illustrate such possibilities in
the present system in the two-parameter $(A-f)$ space, we first
enumerate three standard types of routes to SNA  namely, (1)
Heagy-Hammel ($-$ collision of period doubled torus with its unstable
parent), (2) gradual fractalization ($-$the amplitude or phase
instability), and (3) type--III intermittency ($-$the subharmonic
instability) routes  through which the truncation of torus doubling
bifurcation occurs leading to the birth of SNAs  within the torus
bubble region.

In addition, we identify that in some cross sections of the
$(A-f)$ parameter space, particularly within the torus bubble region,
the period doubling bifurcation phenomenon still persists in the
destroyed torus, eventhough the actual doubling  of the torus
itself has been terminated.  However, we show that {\bf the novel
dynamics involved in this transition is a sudden expansion in the
attractor}. This transition seems to look like the interior crisis which
occurs in low dimensional chaotic systems  \cite {R26}. We  also
demonstrate  the occurrence of SNAs through gradual fractalization and
type--I intermittent nature, during the transition from one-frequency
quasiperiodicity to chaos that exists outside the torus bubble region.

In all our studies the transitions to different SNAs at different 
parts of the border in the $(A-f)$ parameter plane and 
characterization of them is carried out on the
basis of specific quantities such as Lyapunov exponents and their
variance as well as finite-time Lyapunov exponents, dimensions, power
spectral measures and phase sensitivity exponents. Brief details of
these characterizing quantities are given in the Appendix A. In Sec. II  we
describe the phenomenon of the remerging of Feigenbaum tree in the
absence of quasiperiodic forcing in the map (2). The existence of remerging
torus doubling   is pointed out in Sec. III. Various transitions
to SNAs through the truncated doubled torus are demonstrated in Sec. IV. In
particular, the truncation of torus doubling bifurcation and the birth of SNAs
through torus collision, fractalization and type--III intermittent mechanisms
have been explained. Further, a sudden expansion of the  attractor causing the truncation
of torus doubling bifurcation and the genesis of SNA is also demonstrated.
In Sec. V, the transition from one-frequency torus to SNA through
type--I intermittent as well as  fractalization mechanisms is described. 
In Sec. VI the transitions between different SNAs are discuused.
In 
Sec. VII we address the issue of distinguishing among SNAs formed by
different routes through the use of finite-time Lyapunov exponents. Finally,
in Sec. VIII, the results are summarized.

\section{Remerging of Feigenbaum trees in the absence of quasiperiodic forcing}

To start with, we consider the system (2) and numerically iterate it by
varying the values of $A$ and $Q$. For any $Q$ value and low $A$
values, the system (2) exhibits periodic oscillations with period 1T.
As $A$ increases, a bifurcation occurs and the stable period T orbit
transits into a stable period 2T bubble, as shown in Fig.~1(a). For
example, when the value of $Q$ exceeds a certain critical value
$Q$=$-$0.99 for a fixed $A$, $A$=1.5, a transition from period T orbit
to period 2T orbit occurs on increasing $Q$, essentially due to period
doubling bifurcation. Then the period 2T attractor merges and forms a
period T attractor when the value of $Q$  increases to $Q$=0.99 at the
same fixed $A$. At even higher values of $A$, $A$=1.7, the primary
period 2T bubble bifurcates into secondary period 4T bubbles, as shown
in Fig.~1(b). This bubble develops into further bubbles as $A$ gets larger,
until an infinitely branched Feigenbaum tree leading to the onset of
chaos finally appears, as shown in Fig.~1(c) for $A$=1.8.

Bier and Bountis showed that such a remerging of Feigenbaum trees is
quite common in certain models  possessing a kind of reflection
symmetry property coupled with more than one parameter \cite {R25}.
Further, they added that the formation of primary period 2T bubble is
seen to lead to higher order  bubbles and the development of the
associated universal route to chaos in these systems.  It  is also
stated  in the literature that the reversal of period doubling  occurs
when the system possesses a positive  Schwarzian derivative at the
bifurcation point \cite {R27,R28}. This is true for the present case
that we study.  However, there are some counter examples as pointed out
by Nusse and Yorke \cite {R28} to show that the positivity of
Schwarzian derivative  is not a sufficient condition to rule out the
period halving bifurcations.

In the present paper, our aim is to investigate the effect of a
quasiperiodic forcing on the system (2) as given by Eq.(3).
In particular, we point out that with the addition of the quasiperiodic
forcing for a fixed $Q$, the dynamics is dominated by quasiperiodic
attractors and transitions to chaos via strange nonchaotic attractors
(SNAs) along different routes in contrast to the type of attractors 
shown in Fig.~1. For this purpose we also work out a two parameter $(A-f)$
phase diagram (Fig.~2) to identify the changes in the dynamics.

\section{SNAs in the quasiperiodically forced cubic map}

Now we consider the dynamics of the quasiperiodically driven map (3)
and numerically iterate it with the value of the parameter $\omega$
fixed at $\omega = {{\sqrt{5}  - 1} \over 2}$ and by varying  the
values of $A$ and $f$ for different fixed values of $Q$. The results
are then summarized in a suitable two-parameter $(A-f)$ phase diagram
for each fixed value of $Q$. Various dynamical behaviours $-$
quasiperiodic, strange nonchaotic, and chaotic attractors $-$ have been
identified by characterizing the attractors by  quantities such as
Lyapunov exponents and their variance as well as finite-time Lyapunov
exponents, dimensions, power spectral measures and phase sensitivity
exponents  (for  details, see in Appendix A).

In the absence of the external forcing $(f=0)$, from Fig.~1, we can
easily check that for fixed $Q$ and for given $A$ the dynamics
corresponds to periodic or chaotic attractors. For instance, for $Q=0$
and for any values of $A$, the system admits a period-2 solution.
Similarly, for $Q=0.25$ and $A=1.8$, it is a period-4 orbit, while for
$Q=0.5$ and $A=1.8$, it is a chaotic orbit. We now include the effect
of quasiperiodic forcing $(f \ne 0)$ and analyse the dynamics involving
torus, period doubled torus and chaos via SNAs. A very clear picture of
the various types of transitions becomes available for the case $Q=0$
in the region $f \in (-0.8,0.8)$ and $A \in (0.8,2.4)$, while similar
structure arises in a larger region for other values of $Q$.
Consequently, we present in the following  results for $Q=0$ only in
the form of the phase diagram in Fig.~2. The various features indicated
in Fig.~2 are summarised and the dynamical transitions are discussed in
the following.

The general features of the phase diagram fall into a very interesting
pattern.  It can be observed from Fig.~2 that the dynamics is symmetric
about $f=$0. Therefore, in the following we present the details for the
right half of $f=0$ line only. The features are exactly similar in the
left half of the $f=0$ line.  There are two chaotic regions $C1$ and
$C2$. Bordering  these chaotic regions, one has the  regions where the
attractors are strange and nonchaotic. Such SNAs are found to appear in
a large number of regions under various mechanisms, some of which are
marked GF1, GF2, GF3 \& GF4, HH, IC, S1 \& S3.  Besides the strange
nonchaotic  and chaotic attractors in the phase diagram Fig.~2, one can
also observe  different regions where quasiperiodic attractors can  be
found. In Fig.~2, such regions are marked as 1T and 2T, corresponding
to the quasiperiodic attractors of period-1 and period-2 respectively.
Fuller details are given below.

For low $A$ and any $f$ value, the system exhibits quasiperiodic
oscillations denoted by 1T in Fig.~2. On increasing the value of $A$
further, the fascinating phenomenon of torus bubble appears within a
range of values of $f$. To be more specific,the parameter $A$ is for
example fixed at $A$=1.1 and then $f$ is varied. For $f$=$-$0.3, the
attractor is a quasiperiodic one (1T). As $f$ is increased to
$f$=$-$0.18, the attractor undergoes torus-doubling bifurcation and the
corresponding orbit is denoted as 2T in Fig.~2. As $f$ is increased
further, one then expects that the doubled attractor  continues the
doubling sequence as in the case of the generic period-doubling
phenomenon. Instead, in the present case,  the doubled attractor begins
to merge into that of a single attractor at $f$= 0.18, leading to the
formation of a torus bubble reminiscent of period bubbles in low
dimensional systems as in the previous section.  On refixing the
parameter $A$ at higher values, one finds that there are two prominent
regions of chaotic oscillations C1 and C2 as shown in Fig.~2. The
chaotic region C1 exists  outside the torus bubble region. That is, it
essentially occurs for larger A values, $A>$1.2 and  $f>$0.6.  On the
other hand, the region C2 emerges within the torus bubble region. That
is, it appears predominantly  for even larger values of $A$, $A>$1.549
and $f$ lying between, $-$0.8$<f<$0.8.  We have identified two
interesting dynamical transitions from one frequency quasiperiodicity
to chaos via SNAs outside the torus bubble region.  They are (1)
gradual fractalization of the torus leading to creation of SNA (GF4),
and (2) type--I intermittent route leading to the birth of SNA (S1).
On the other hand there exist atleast four types of transitions to
chaos via SNAs within the torus bubble where the doubling of torus is
interrupted, namely (1) Heagy-Hammel (HH), (2) fractalization
(GF1, GF2 \& GF3), (3) type--III intermittent (S3) and (4) doubling of
destroyed tori routes through which the torus doubling  bifurcation is
truncated and the birth of strange nonchaotic attractor takes place.
The details for each of the regions are given in the following sections.

\section{Dynamics within the torus bubble} 

In this section, we will describe each one of the four types of transition
to chaos via 
SNAs within the torus bubble region in detail.

\subsection{Heagy-Hammel route} 

The first of the routes which we encounter is the Heagy-Hammel route in
which a period$-$ 2$^n$ torus gets wrinkled and upon collision with its
unstable parent period$-$ 2$^{n-1}$ torus bifurcates into an SNA \cite
{R8}.  Such a route has been identified in the region C2 within the
range of $A$ values, 1.549$<A<$2.183, and $f$ values, 0.39$<f<$0.8.
That is the doubling bifurcation is truncated due to the collision of
the  doubled torus with its unstable parent on increasing the value of
A, in the range 1.549$<A<$2.183, for a fixed $f$ value (0.39$<f<$0.8).
It is denoted as HH in Fig.~2. For example, let us fix the parameter
$f$ at $f$=0.7 and vary $A$.  For $A$=1.8, the attractor is a
quasiperiodic one, as denoted by 1T in Fig.~2. As $A$ is increased to
$A$=1.876, the attractor undergoes torus doubling bifurcation and the
corresponding periodic orbit is denoted as 2T in Fig.~2.  In the
generic case, the period doubling occurs in an infinite sequence until
the accumulation point is reached, beyond which chaotic behaviour
appears. However, with tori, in the present case, the truncation of the
torus doubling begins when the two strands of the 2T attractor become
extremely wrinkled. For example, when the value of $A$ is increased to
$A$=1.8868, the attractor becomes wrinkled as shown in Fig.~3(a). At
this transition, the strands are seen to come closer to the unstable
period 1T orbit and lose their continuities when the strands of torus
doubled orbit collide with unstable parent  and ultimately result in a
fractal attractor as shown in Fig.~3(b) when $A$ is increased to
$A$=1.88697.  At such a value, the attractor, Fig.~3(b), possesses a
geometrically strange property but does not exhibit any sensitivity to
initial conditions (the maximal Lyapunov exponent is negative as seen
in Fig.~4(a)) and so it is indeed a strange nonchaotic attractor.  At
this transition, the two branches of the wrinkled attractor collide and
form a one  band SNA. This kind of transition is similar to the
attractor merging crisis occuring in chaotic systems \cite {R26}. As
$A$ is increased further to $A$=1.8878, the attractor has eventually a
positive Lyapunov exponent and hence it corresponds to a chaotic
attractor (C2).

Now we examine the Lyapunov exponent for the transition from torus to
SNA.  Fig.~4(a) is a plot of the maximal Lyapunov exponent as a
function of $A$ for  $f$=0.7. When we examine this in a
sufficiently small neighbourhood of the critical value $A_{HH}$=1.88697,
the transition is clearly  revealed by the Lyapunov exponent which
varies smoothly in the torus region ($A<A_{HH}$) while it varies
irregularly in the SNA region ($A>A_{HH}$).  It is also possible to
identify this transition point by examining the variance of the Lyapunov
exponent, as shown in Fig.~4(b) in which the fluctuation is small in the
torus region while it is large in the SNA region.

In addition, in order to distinguish the quasiperiodic attractor and
the strange nonchaotic attractor, we may examine the attractor with
reference to the phase $\theta$ of the external force. The details of
this analysis are given in Appendix A. From Eq.(A.4), one infers that
the function $\Gamma_N$ grows infinitely for a SNA with some relation
such as $\Gamma_N = N^{\mu}$, where $\mu$ is a positive quantity which
characterises the SNA and we may call it the phase sensitivity
exponent. For the present case, it is $\mu$=0.98. However, in the case
of a chaotic attractor, it grows  exponentially with N (see Fig.
4(c)).

\subsection{Fractalization route} 

The second  of the routes is the gradual fractalization route where a
torus gets increasingly wrinkled and then transits to a SNA without
interaction (as against the previous case) with a nearby unstable orbit
as we change the system parameter. In this route, a period$-$2$^n$
torus becomes wrinkled and then the wrinkled attractor gradually loses
its smoothness and forms a 2$^n-$band SNA as we change   the system
parameter. Such a phenomenon has been identified in the present system
in three different regions indicated as GF1, GF2 and GF3 in Fig. 2.  To
exemplify this nature of  transition, we fix the parameter $f$ at
$f$=0.1 and vary $A$ in the GF3 region.  For $A$=1.0, the system
exhibits quasiperiodic oscillation of period-1T. The attractor
undergoes a torus doubling bifurcation as $A$ is increased to
$A$=1.06.  Increasing the $A$ value further, a second period doubling
of the doubled torus does not take place. Instead, oscillations of the
doubled torus in the amplitude direction starts to appear at $A$=2.165
as shown in Fig.~5(a).  As $A$ is increased further to $A$=2.167, the
oscillatory behaviour of the torus approaches a fractal nature
gradually. At such values, the nature of the attractor is strange [see
Fig.~5(b)] eventhough the largest Lyapunov exponent in Fig.~6(a)
remains negative.  Such a phenomenon is essentially a gradual
fractalization of the doubled torus as was shown by Nishikawa and
Kaneko in their route to chaos via SNA \cite {R9}. In this route, there
is no collision involved among the orbits and therefore the Lyapunov
exponent increases only slowly as shown in Fig.~6(a) and there are no
significant changes in its variance [see  Fig.~ 6(b)].  Further, the
phase sensitivity function $\Gamma_N$ grows unboundedly with  the
power-law relation $\Gamma_N= N^{\mu}$,  $\mu$=0.83 in the SNA region,
while they are bounded in the torus region [see Fig. 6(c)]. At even
higher values of $A$, $A$=2.17, the system exhibits chaotic
oscillations (C2). The quantity $\Gamma_N$ grows exponentially with N
for the chaotic attractors.

\subsection{Type III intermittent route} 

The third one of the routes that is predominant in this system within
the torus doubled region is an intermittent route in which the torus
doubling bifurcation is tamed due to subharmonic bifurcations leading
to the creation of SNA. Such a phenomenon has been identified  within
the range of $f$ values 0.33$<f<$0.41 and on increasing the value $A$,
1.81$<A<$2.18, for a fixed $f$.  To illustrate this transition, let us
fix the parameter $f$ at $f$=0.35 and vary $A$. For $A$=1.0, the
attractor is quasiperiodic attractor. As $A$ is increased to $A$=1.28,
the attractor undergoes torus doubling bifurcation. On increasing the
value of $A$ further, $A$=2.13, the attractor starts to wrinkle. On
further increase of $A$ value to $A$=2.135, the attractor becomes
extremely wrinkled and has several sharp bends, as shown in Fig.
7(a).   It has been observed in lower dimensional chaotic systems
\cite{R28,R29} that when the system undergoes subcritical period
doubling bifurcation, the dynamical behaviour exhibits type III
intermittent motion. In a similar manner,one finds
 that the wrinkled attractor undergoes a quasiperiodic analogue of
subcritical period doubling bifurcation, on increasing the value of $A$
further to $A$=2.14. The corresponding intermittent motion is shown in
Fig.~7(b).  The emergence of such intermittent dynamical behaviour has
been found in different continuous systems by the present authors and
their collaborators through the intermittent route  to chaos via SNA
\cite {R12d} in which it was shown that during the transition from
torus doubled attractor to SNA, a growth of subharmonic amplitude
begins together with a decrease in the size of the fundamental
amplitude.  At the critical parameter value, the intermittent attractor
loses its smoothness and becomes strange.  The attractor shown in
Fig.~7(b) is nothing but a strange nonchaotic one as the Lyapunov
exponent turns out to be negative [Fig.~8(a)].  When examining the
Lyapunov exponent at this transition, it has been observed in Fig.~8(a)
that the Lyapunov exponent shows an abrupt change with a power-law
dependence on the parameter on the SNA side of the transition and the
variance shows a remarkable and abrupt increase at the transition point
as shown in Fig.~8(b). Further, the phase sensitivity function
$\Gamma_N$ is bounded for the torus region while it is unboundedly
changing with a power-law variation with $N$ for SNA region [Fig.~8(c)]
with $\mu =0.85$.  On increasing the value of $A$ further to $A$=2.153,
we find the emergence of chaotic attractor (C2) where the quantity
$\Gamma_N$ grows exponentially with N [Fig.~8(c)].

In the HH case, the points on the SNA are   distributed over the entire
region enclosed by the wrinkled bounding torus, while  in the GF case
the points on the SNA are distributed mainly on the boundary of the
torus.  Interestingly in the present case shown in Fig.~7(b), most of
the points of the SNA remain within the wrinkled torus with sporadic
large deviations.  The dynamics at this transition obviously involves a
kind of   intermittency. Such an intermittency transition could be
characterised by the scaling behaviour. The laminar phase in this case
is the torus while the burst phase is the nonchaotic attractor.  In
order to calculate the associated scaling constant, we coevolve the
trajectories for two different values of $A$, namely $A_c$ and another
nearby value of $A_c$, while keeping identical initial conditions
($x_i$,$\theta_i$) as well as the same parameter value $f$. As the
angular coordinate $\theta_i$ remains identical, the difference in
x$_i$ allows one to compute the average laminar length between the
bursts and it fits with  the scaling form $$ <l> = (A_c-A)^{-\alpha}.
\eqno(4) $$ The numerical value obtained for the attractor shown in
Fig. 7(b) is $\alpha
 \sim$ 1.1 [see Fig.~9(a)]. To confirm further that the SNA attractor
[Fig.~7(b)] is associated with intermittent dynamics, we plot the
frequency of laminar periods of duration $\tau$, namely N($\tau$) in
Fig.~9(b).  It obeys the scaling \cite {R30} law, $$ N(\tau) \sim \left
\{ {\exp (-4 \epsilon \tau) \over [1- \exp(-4 \epsilon \tau)]
 } \right \}  ^{0.5}. \eqno(5) $$ We find that
$\epsilon$=0.007$\pm$0.0002 to give a best fit for the present data.
These characteristic studies  suggest that the intermittency is of type
III as discussed by Pomeau and Manneville in low dimensional systems
\cite {R29,R30}.

\subsection{Crisis-induced intermittency} 

In the previous subsections, we have seen that the period doubling
bifurcation of a torus has been truncated by the destruction of it
leading to the emergence of an SNA in certain regions of the $(A-f)$
parameter space.  Further, we observe that in the present system  in
some cross sections of the $(A-f)$ parameter space the period doubling
phenomenon still persists in the destroyed torus eventhough the actual
doubling phenomenon has been truncated. But  in the present case, it is
observed that the doubling of destroyed torus involves a kind of sudden
widening of the attractor similar to the crisis phenomenon that occurs
in chaotic systems. Such a phenomenon has been observed in the present
model in a range of $f$ values, 0.13$<f<$0.24, and for a narrow range of
$A$ values, 2.12$<A<$2.14. It is denoted as IC in Fig.~2. For
example, let us choose $f$=0.2 and vary the value of $A$. For $A$=0.8,
the attractor is a quasiperiodic  one.  As $A$ is increased to
$A$=1.18, the system undergoes torus doubling bifurcation. On
increasing the value of $A$ further to $A$=2.138, the attractor begins
to wrinkle  as shown in Fig. 10(a). On increasing the value of $A$
further, say, to $A$=2.1387 the  wrinkled attractor undergoes torus doubling
bifurcation and the corresponding orbit is shown in Fig.~10(b). It is
also seen from Figs.~10(b) \&(c) that when A is slightly larger than
$A_{IC}=2.1387$ the orbit on the attractor spends long stretches of
time in the region to which the attractor was confined before the
crisis. At the end of these long stretches the orbit bursts out of the
old region and bounces around in the new  region made available to it
by the crisis. It then returns to the old region for another stretch of
time, followed by a burst and so on.  This  kind of widening of
attractor  usually occurs in the chaotic systems at a crisis \cite
{R26}. However, in the present case, we have shown such a possibility
in a quasiperiodically forced system  which also tames the growth of
torus doubled cascade, giving birth to SNAs. The variation of the
Lyapunov exponent at such transition does not follow a uniform pattern
in   contrast to the case of low-dimensional chaotic systems exhibiting
crisis phenomenon [see Figs.~11 (a) \& (b)]. In addition, the phase
sensitivity function  $\Gamma_N$ grows with $N$ with a kind power-law
relations for SNA while it is bounded  for the torus regions [see
Fig.~11 (c)]. On further increase of  the value of $A$ to $A$=2.143,
the system exhibits chaotic oscillations (C2) and the quantity
$\Gamma_N$ grows exponentially with $N$.

\section{Dynamics outside the torus bubble}

Two additional interesting transitions take place outside the torus
bubble region, namely (1) gradual fractalization of the torus leading
to the creation of SNA, and (2) type--I intermittent route leading to the
birth of SNA.  The details are as follows.

\subsection{Fractalization route}

The  first one is the gradual fractalization route which is the same as
studied in the previous subsection IV B, but the only difference now is
that here a transition from a one-frequency torus (1T) to chaos via
SNA is realized through gradual fractalization process instead of the
transition from the 2T torus discussed above. Such a phenomenon is
identified in the lower side of the C1 region (GF4).  Specifically, within
the region of $f$,0.58$<f<$0.8, on increasing the value of $A$  to the region
1.211$<A<$1.569 for a
fixed $f$ the SNA is created through gradual fractalization route. At
first, a transition from one-frequency torus to a wrinkled attractor
takes place on increasing $A$, $A$=1.25 for $f$=0.7 as shown in Fig.12(a). The
wrinkled attractor loses its continuity considerably as $A$ is
increased further and then finally ends up with fractal phenomenon at
$A$=1.265 (see fig. 12(b)).  It is very obvious from these transitions
that the torus  gradually loses its smoothness and ultimately
approaches the fractal behaviour via SNA before the onset of chaos as
the parameter $A$ increases to $A$=1.3. In addition, it is observed that
there is no apparent interaction among the orbits. The property has been confirmed 
through the calculation of maximal Lyapunov exponent, its variance and phase sensitivity
studies as in the period-doubled regions.

\subsection{Type--I intermittent route} 

A different type of intermittent route, namely type--I \cite {R14}, via
SNA is also observed at the upper region of C1.  Within the range of
values $f$,0.58$<f<$0.8, on increasing the value of $A$ in the range 1.5
$<A<$2.0, a transition from the chaotic attractor (C1) to SNA takes
place at first and  then the SNA is eventually replaced by a
one-frequency quasiperiodic orbit through a quasiperiodic analogue of
saddle-node bifurcation. At this transition, the dynamics is found to
be again intermittent but of different type.

To understand more about this phenomenon, let us consider the specific
parameter value $f$= 0.7 and vary $A$. For $A$=1.80165, the attractor is a
chaotic one (C1). As $A$ is increased to $A$=1.801685, the chaotic
attractor transits to SNA as shown in Fig. 13(a). On increasing the value
of $A$ further, an intermittent transition from the SNA  to a torus as
shown in Fig. 13(b) occurs at $A$=1.8017 . At this transition,  abrupt
changes in the Lyapunov exponent as well as its variance shows the
characteristic signature of intermittent route (indicated in Fig. 14 (a)
$\&$ (b)) to SNA as in the type III case. In addition, again
the quantity $\Gamma_N$ grows with $N$ with a power law relation for SNA while
it is bounded  for the torus regions (see Fig. 14 c). However, in the chaotic
region,  the quantity $\Gamma_N$ grows exponentially
with N. 

Further, the plots of laminar
length $<l>$ as a function of the derived bifurcation parameter
$\epsilon =A-A_c$, where A$_c$ is a critical parameter for the
occurrence of intermittent transition, for this attractor reveals a
power-law relationship of the form 
$$
 <l> =  \epsilon^{-\beta}, \eqno(6) 
$$
with
an estimated value of $\beta \sim $ 0.53 (fig. 15(a)). 
Also, the plot
between the number of laminar periods N($\tau$) and the  period length
$\tau$ [shown in Fig.~15(b)] indicates  that after an initial  steep decay
there is an increase to a large value of
N($\tau$). It also obeys the relation  
$$
 N(\tau) \sim { \epsilon \over 2c} \left \{
\tau+ \tan \left [ \arctan \left ({c \over \sqrt {\epsilon \over
u}}\right ) - \tau \sqrt{\epsilon u} \right ] - \arctan \left ({c \over
\sqrt { \epsilon \over u}} \right ) \tau - \tau^2 \sqrt{\epsilon \over
u}
\right \}, \eqno(7)
$$
 where $c$ is the maximum value of $x(t)$, $u=0.9$
and $\epsilon=0.0005 \pm 0.00003.$
The above  analysis  
confirms that such an attractor is associated with the standard
intermittent dynamics of type I described in ref. \cite {R29,R30}.

\section{Transition between different SNAs}

In the previous sections, we have enumerated the several ways through which
SNAs are created from torus attractors.  One might  observe from the $(A-f)$
phase diagram, Fig.~2, that there are several regions where transitions from
one type of SNA to another type  occurs along the borders separating 
quasiperiodic and chaotic attractors. Particularly,
transitions occur between GF3 and IC, IC and GF2, GF2 and
S3, S3 and GF1, GF1 and HH, and,  S1 and GF4. On closer scrutiny, 
we find that there exists
a very narrow range of parameters between different regions of SNAs  where
chaotic motion occurs. That is the SNA of one region transits to chaos before 
exhibiting 
a different nature of SNA in the next region. However, we refrain from giving
finer details as they do not seem to be of significance.

\section{\protect\smallskip Signatures of finite time Lyapunov
exponents at the transition to Different SNAs}

Recently, it has been noted by Prasad, Mehra and Ramaswamy \cite {R14}
that a typical trajectory on a SNA\ actually possesses positive
Lyapunov exponents in finite time intervals, although the asymptotic
exponent is negative. As a consequence, one observes the different
characteristics of the SNAs born through different mechanisms through a
study of the differences in the distribution of finite-time exponents
P(N,$\lambda $) \cite{R14}.  For each of the cases, the distribution
can be obtained by taking a long trajectory and dividing into segments
of length N, from which the local Lyapunov exponent can be calculated.
In the limit of large N, this distribution will collapse to a $\delta$
function $P(N,\lambda) \rightarrow \delta (\Delta - \lambda)$.  The
deviations from$-$ and the approach to $-$ the limit can be very
different for SNAs created through different mechanisms.  Figs.~16
illustrate the distributions for $P(50,\delta)$ across the five
different transitions discussed in the present study.  A common feature
of these cases is that $P(N,\lambda)$ is strongly peaked about the
Lyapunov exponent when the attractor is a torus, but on the SNA, the
distribution picks up a tail which extends into the local Lyapunov
exponent $\lambda > $0 region. This tail is directly correlated with
enhanced fluctuation in the Lyapunov exponent on SNAs.  On the
fractalised SNA and doubling of SNA, the distribution shifts
continuously to larger Lyapunov exponents, but the shape remains the
same for torus regions as well as SNA regions while on the HH and
intermittent SNA, the actual shapes of the distribution on the torus
and the SNA are very different. One remarkable feature of intermittent
SNAs is that the positive  tail in the distribution decays very
slowly.

To quantify further on the distribution of finite-time Lyapunov
exponents, let us consider, for example, the fraction of exponents lying
above $\lambda$=0, $F_{+}(N)$ vs $N$ for the different SNAs. It has
been found that  except for the intermittent SNA (both for type III and
I), for which $F_{+}(N) \sim N^{-\beta}$, this quantity decays
exponentially , $F_{+}(N) \sim exp (- \gamma N)$ for all other
transitions, with the exponents $\beta$ and $\gamma$ dependent strongly
on the parameters of the system. For the specific SNAs corresponding to
the parameters reported in the previous section these quantities take
the following values: For type III and type I, the $\beta$ values are
0.38 and 0.71. However, for HH, GF and IC, the $\beta$ and $\alpha$
values are 0.27 \& 0.32, 0.31 \& 017 and 0.24 \& 0.34, respectively.

\section{Conclusion} 

In this paper we have described the creation of SNAs through various
routes and mechanisms in a protypical example, namely the
quasiperiodically driven cubic map. These are summarised in Table I.
Torus doubling bifurcations are not mandatory for the creation of SNAs.
However, they are merely a convenient agent in setting the stage for
the appearance of SNAs.  There are atleast four different mechanisms,
namely Heagy-Hammel, gradual fractilazation, type--III and and
crisis-induced intermittency through which the truncation of torus
doubling and the birth of SNAs occur.  The truncation of torus doubling
and the genesis of SNA through crisis-induced intermittency is a new
one and also it is entirely different from the interior crisis
mechanism for appearence of SNAs as found by Witt {\it et al.} \cite{R33} We
have further observed atleast two different ways, namely type--I
intermittency and gradual fractalization, through which SNAs are  formed
when a transition from one-frequency torus to chaos takes place.  All
these phenomena have been identified in a two parameter $(A-f)$ phase
diagram. 
To distinguish among the different mechanisms through which SNAs are
born, we have examined the manner in which the maximal Lyapunov
exponent and its variance change as a function of the parameters. In
addition, we have also examined the distribution of local Lyapunov
exponents and found  that on different SNAs they have different
characteristics. The analysis confirms that in the intermittent SNAs,
the signature of the transition is a discontinuous change in both the
maximal Lyapunov exponent and  in the variance. The chaotic component
on the intermittent SNA is long lived.  As a consequence, a slow
positive tail in the P(N,$\lambda$) and a resulting power-law decay for
F$_+$(N) can be identified.  For the other SNAs, the resulting
exponential decay for F$_+$(N) has been identified.

\section*{Acknowledgment} 

This work forms of a Department of Science and Technology, Government
of India research project. A. V. wishes to acknowledge Council of
Scientific and Industrial Research, Government of India, for financial
support.

\appendix
\section{Characterizations of SNA}
\subsection{ Phase sensitivity exponent}
In order to distinguish the smooth and the fractal torus (SNA and chaos),  we 
examine the attractor with reference to the phase $\theta$ of the external force.
Eventhough, no exponential divergence of orbits exists for both the smooth 
and the fractal torus (SNA), they are different from each other in terms of the phase
sensitivity. Pikovsky and  Feudal \cite {R16a,R17} have shown that how two points on the SNA
which have close $\theta$ values separate from each other by introducing the following phase
sensitivity exponent: To appreciate this, we note that the absolute value of the first
derivative of the orbit$| {\partial x_n \over \partial \theta_n} | $ fluctuates with time
and some times has large bursts.To see this, one can proceed as follows. An aribitarary large burst can appear when the 
system is iterated for infinite time steps. By differentiating (3), one
obtains
$${\partial x_{n+1} \over \partial {\theta}} = -2\pi Q sin(2 \pi {\theta})-(A-3x_n^{2})
{\partial x_n \over \partial {\theta}}. \eqno(A.1) $$
So, starting from a suitable initial derivative  $| {\partial x_0 \over \partial {\theta}} | $,
one can obtain derivatives at all points of the trajectory
$${\partial x_{N} \over \partial  {\theta}} =S_N= \sum_{k=1}^{N} \left \{-2\pi Q sin(2 \pi  {\theta}_{k-1})
\left [\prod_{i=0}^{N-k-1}-(A-3x_{k+i}^{2}) \right ] \right \}
+ \prod_{i=0}^{N-1} \left (-(A-3x_{i}^2){\partial x_0 \over \partial {\theta}} \right ), \eqno(A.2) $$
with the condition that for $k=N$,  
$$\prod_{i=0}^{N-k-1}\left [-(A-3x_{k+i}^{2}) \right ]=1. \nonumber$$ Naturally, 
in the case of a smooth attractor   if one iterates (A.1) and (A.2)  starting from 
aribitrary values of $x$ and $ {\partial x \over \partial \theta}  $ and for large $N$,  they 
converge to the attractor and its derivative, respectively. Thus,  partial sums
$S_N$ computed from (A.2) are bounded by the maximum derivative $ {\partial x \over \partial \theta}$
along the attractor. But in the case of fractal attractor, the attractor is nonsmooth and the 
derivative $ {\partial x \over \partial \theta}$ does not exist, so the consideration above is no longer
valid. These can be illustrated by calculating the partial sums $S_N$ as given by (A.2). It has been found \cite{R16a,R17}
 that  the behviour of the sum can be 
very intermittent. The key observation is that these sums are quite large and practically unbounded. Hence, 
we plot the  maximum of $|S_N|$,  
$$\gamma_N(x,\theta)=max|S_N|. \eqno (A.3) $$
The value of $\gamma_N$ grows with $N$, which means that arbitararily large values of $|S_N|$ apppear. From this 
it follows immediately that the attractor can not have finite derivative with respect to the external phase if the
attractor is nonsmooth. Consequently, the assumption of a finite derivative is inconsistent with the relation (A.2),
where the second term on the RHS is exponentially small and the first term on the RHS can be arbitrary large. Thus,
calculating the partial sums (A.2), we can distinguish strange (sums are unbounded) and nonstrange (sums are bounded)
attractor.

The growth rate of the partial sums with time represents a degree of
strangeness of the attractor, and can be used as a quantitaive
characteristic of SNAs. For this purpose, we require a quantity which
is independent of a particular trajectory while it represents average
properties of the attractor. The appropriate quantity seems \cite{R16a,R17} to be the
minimum value of $\gamma_N(x,\theta)$ with respect to randomly chosen
initial points $(x,\theta)$:  
$$\Gamma_N=min \gamma_N(x,\theta).  \eqno
(A.4) $$
 It allows a more reliable inference about whether the attractor is
 nonsmooth.  One can also infer from (A.4) that $\Gamma_N$ grows
infinitely for a SNA with some relations as $\Gamma_N = N^{\mu}$,  where
$\mu$ is a postive quantity which characterises the SNA and we call it
the phase sensitivity exponent. However, in the case of a chaotic
attractor, it grows exponentially with $N$.

\subsection{Variation of finite--time Lyapunov exponents}

As the finite--time or  local Lyapunov exponents ($\lambda_i$, $i=1,2,\cdots,M$)
depend on the initial conditions,
it will be relevant to consider the variance of the average Lyapunov
exponent $\Lambda $ about the $\lambda_i$'s, $
i=1,2,\cdots,M$.It is  defined \cite{R12d,R14,R26,R32} as 
$$
\sigma=\frac {1}{M} \sum_{i=1}^{M} (\Lambda-\lambda_i(N))^2. \eqno(A.5) $$

In all our numerical calculations, we take $N=50$ and $M \cong 10^5$.

The variation of the local Lyapunov exponents in a fixed time
interval $t$ can also be discussed by
examining  the probability distributions $P(t,\lambda)$ for the exponents.
In fact, $P(t,\lambda)$ corresponds to counting  the normalized number of times any one of the 
$\lambda$ appears for fixed time $t$. That is,
the distribution of local Lyapunov exponents, which is a stationary quantity,
is defined as \cite{R26}
$$
P(t, \lambda) d\lambda \equiv \mbox{ Probability that}
~\lambda(t)
\mbox{~takes a value between} ~~\lambda \mbox{~and~} \lambda + d\lambda. \eqno(A.6)
\label{prob}
$$
This is particularly useful in describing the structure and dynamics of
nonuniform attractors. In the asymptotic limit $t
\to \infty$, this distribution will collapse to a $\delta$ function,
$$P(t,\lambda) \to \delta (\Lambda - \lambda).$$ 
The deviations from this limit for finite
times, and the asymptotics, namely the approach to the limit can be
very revealing of the underlying dynamics \cite{R14}.

One can also 
calculate the arithmetic mean of all the  distributions and obtain the variance
of the Lyapunov exponent $\Lambda$ as
$$
\sigma = \int_{ - \infty}^{\infty} (\Lambda - \lambda)^2 P(t,\lambda)
d \lambda.  \eqno(A.7)
$$
Dividing the total length of the orbit into $M$ bins as before and defining 
the local Lyapunov exponents as $\lambda_i$, replacing $P(t,\lambda_i)$ by $\frac {\delta(\Lambda-\lambda_i)}
{M}$, the above equation of variance goes over to the 
form given by Eq.~(A.5).

\subsection{ Power spectrum analysis}

To quantify the changes in the power spectrum (obtained using Fast Fourier Transform (FFT)
technique), one can compute the so called spectral distribution function $N ( \sigma) $,
defined to be the number of peaks in the Fourier amplitude spectrum larger than
some value say $ \sigma $. Scaling relations have been predicted for $N ( \sigma) $
in the case of two and three frequency quasiperiodic attractors and strange nonchaotic attractors.
These scaling relations are $N(\sigma) \sim \ln {1 \over \sigma}$,$N(\sigma) \sim \ln^2 {\sigma},$
and $N(\sigma) \sim \sigma^{-\beta} $, respectively, corresponding  to the two, three frequency
quasiperiodic and strange nonchaotic attractors. In the work of Romeiras and Ott \cite{R12},
 the power law exponent was found empirically to lie with in the range
$1 < \beta < 2$ for the strange nonchaotic attractor. 

\subsection{ Dimensions}
To quantify geometric properties of attractors, several methods have been used to
compute the dimension of the attractors. Among them,  we have used  the 
correlation dimension (introduced by Grassberger and Procaccia \cite{R31}) in our present
study, which may be computed
from the correlation function C(R) defined as
$$ C(R) =\lim_{N \to \infty} \left [ {1 \over N^2} \sum_{i,j=1}^N H(R- \mid x_i -x_j \mid) \right ], \nonumber $$
\noindent
where $x_i$ and $x_j$ are points on the attractor, H(y) is the Heaviside function
(1 if y $\geq$ 0 and 0 if y $ < $ 0), N is the number of points randomly chosen
from the entire data set. The Heaviside function simply counts the number of points
within the radius R of the point denoted by $x_i$ and C(R) gives the average fraction of 
points. Now the correlation dimension is defined by the variations of C(R) with R:
$$ C(R) \sim R^d    \hskip 20pt \mbox{as} \hskip 20pt  R \rightarrow 0. \nonumber $$
Therefore the correlation dimension (d) is the slope of a graph of log C(R) versus log R.
Once one obtains the  dimensions of the attractors, it will be easy   to quantify strange
property of the attractors. In all our studies, we have verified that the SNAs have noninteger
correlation dimensions.

%\end{document}

%\end{document}
%\newpage 
\begin{figure} 
\centerline{\epsfig {figure=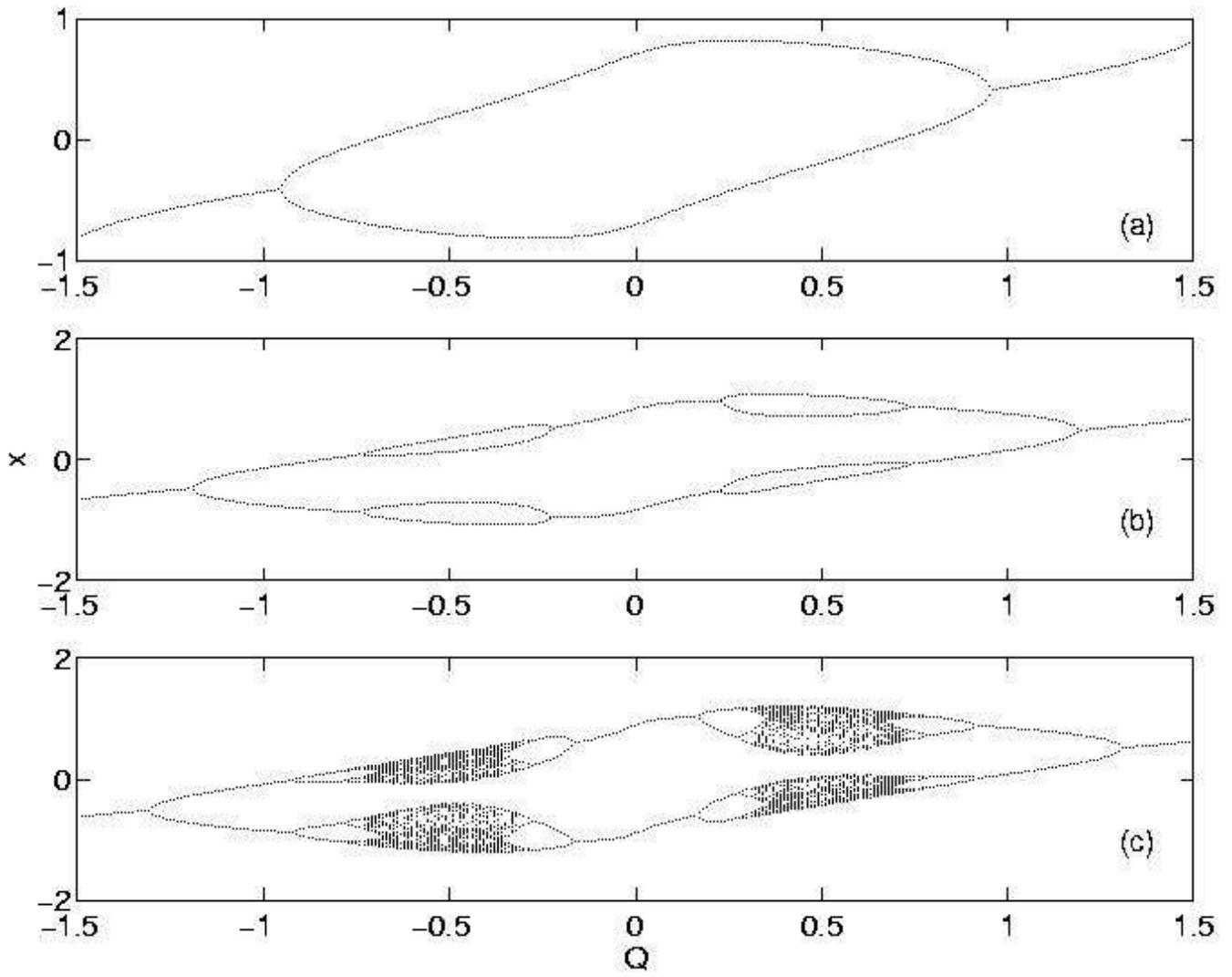,width=\linewidth}} 
\caption{Bifurcation diagram for the map
(2) in the $(x,Q)$ plane: (a) The primary bubble at $A$=1.5, (b) Period-2 and period-4
bubbles at $A$=1.7, (c)  Period doubling route to chaos and inverse period doubling at $A$=1.8.}
\end{figure}
%\end{document}
\begin{figure} 
\centerline{\epsfig {figure=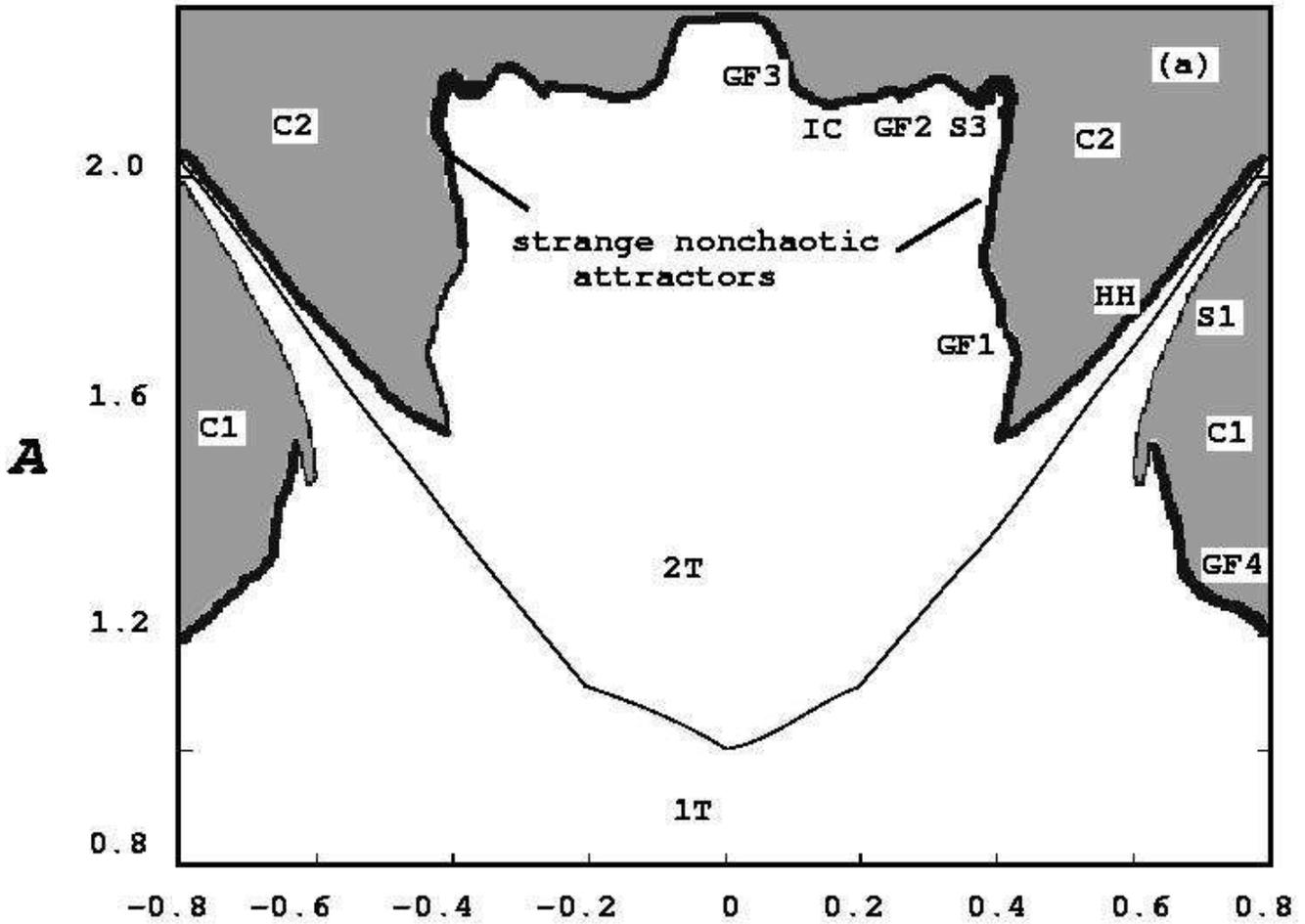,width=\linewidth}}
\caption{Phase diagram for  quasiperiodically forced cubic map, Eq.~(3), 
in the$(A-f)$
parameter space for $Q=0$: Here 1T and 2T correspond to torus of period one and
of two respectively. GF1, GF2, GF3 and GF4 correspond to the regions
where the process of gradual fractalization of torus occurs. HH
represents the regions where SNA is created through the Heagy-Hammel
route.  S1 and S3 denote regions where the SNA appears through Type-I
and Type-III intermittencies, respectively. IC denotes region where
the SNA is created through  crisis-induced intermittency. C1 and  C2  correspond to
chaotic attractors. 
}
\end{figure}
\begin{figure}
\centerline{\epsfig {figure=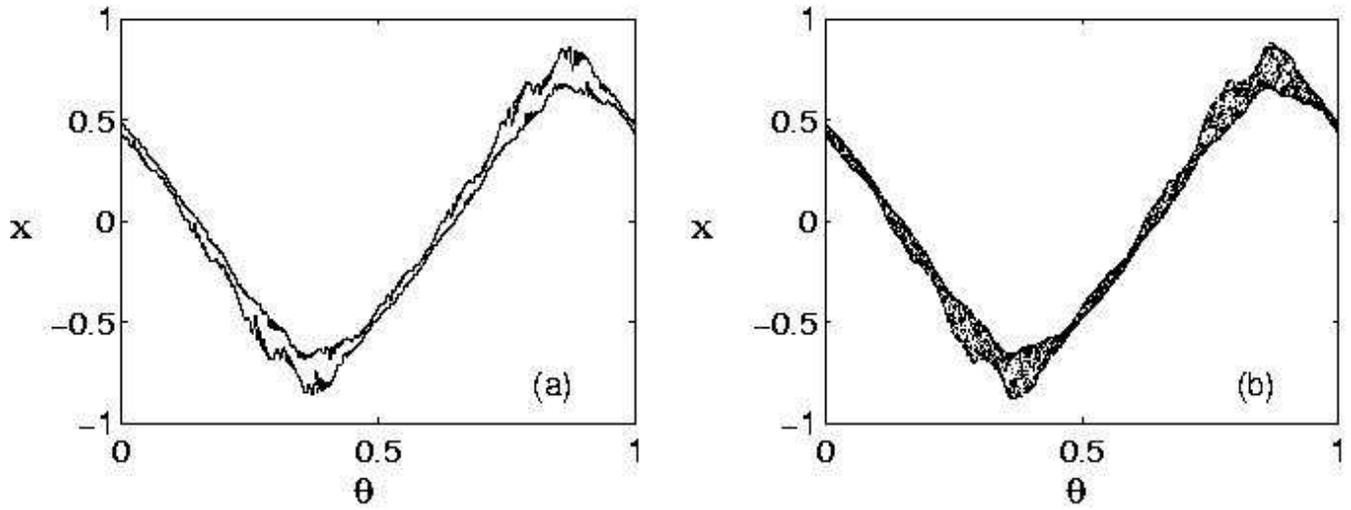, width=\linewidth}}
\caption{Projection of the   attractors of Eqs.~(3) for
$f$=0.7 in the $(x,\theta)$ plane  indicating
the transition from  quasiperiodic attractor to chaotic  attractor via SNA through
Heagy- Hammel mechanism:
(a) wrinkled attractor (period 2T)  for $A$ = 1.8868; (b)
SNA at $A$=1.88697.}
\end{figure}
\begin{figure}
%\newpage
\centerline{\epsfig {figure=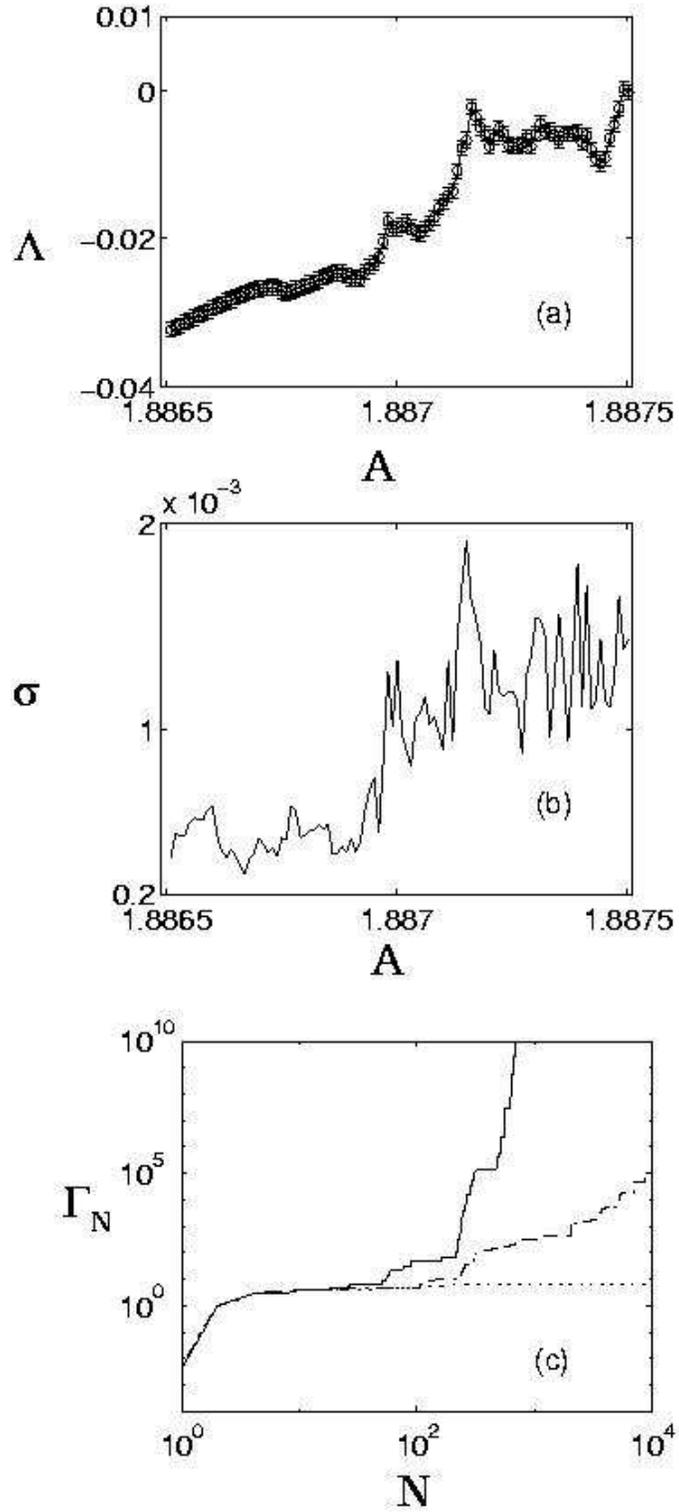, width=0.5\linewidth}}
\caption{Transition from doubled torus to SNA through Heagy- Hammel
mechanism in the region HH: (a) the behaviour of the
Lyapunov exponent ($\Lambda$); (b) the variance ($\sigma$); (c) plot of phase
sensitivity function $\Gamma_N$ vs $N$ (dotted line corresponds to torus for $A$=1.83,
dashed line belongs to SNA for $A$=1.88697 and solid line represents  chaos for $A$=1.8878).}
\end{figure}
\begin{figure}
\centerline{\epsfig {figure=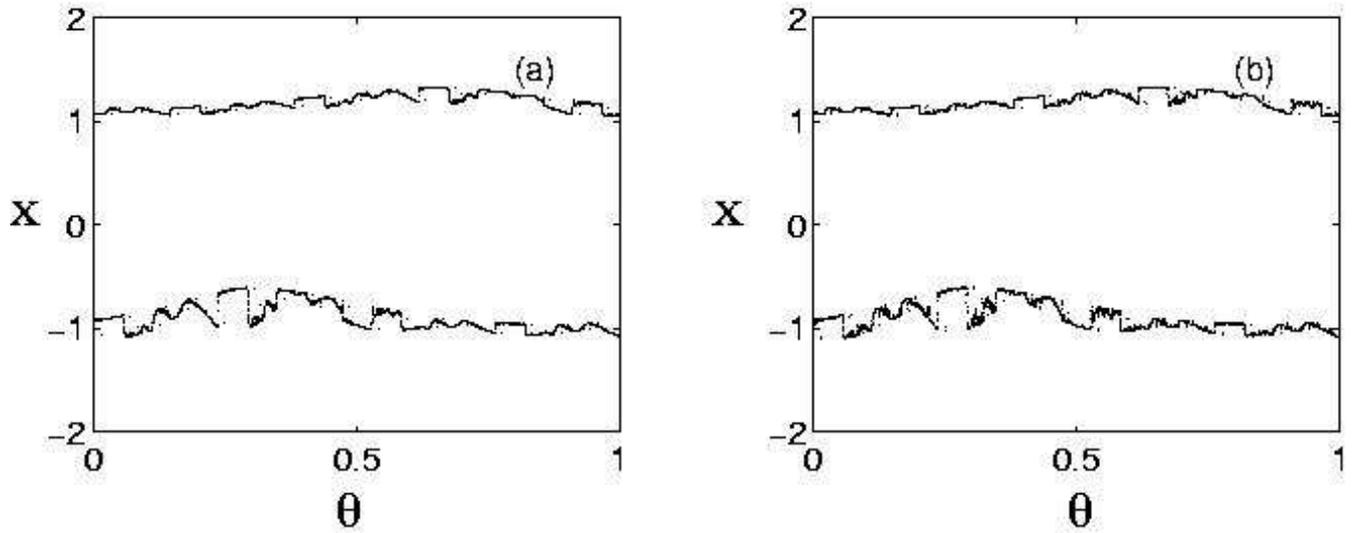, width=\linewidth}}
\caption{Projection of the  attractors of Eqs.~(3) for
$f$=0.1 in the $(x,\theta)$ plane  indicating
the transition from  quasiperiodic attractor to chaotic attractor  via SNA through
gradual fractalization mechanism:
(a) wrinkled attractor (period 2T)  for $A$ = 2.165; (b)
SNA at $A$=2.167.}
\end{figure}
%\newpage
\begin{figure}
\centerline{\epsfig {figure=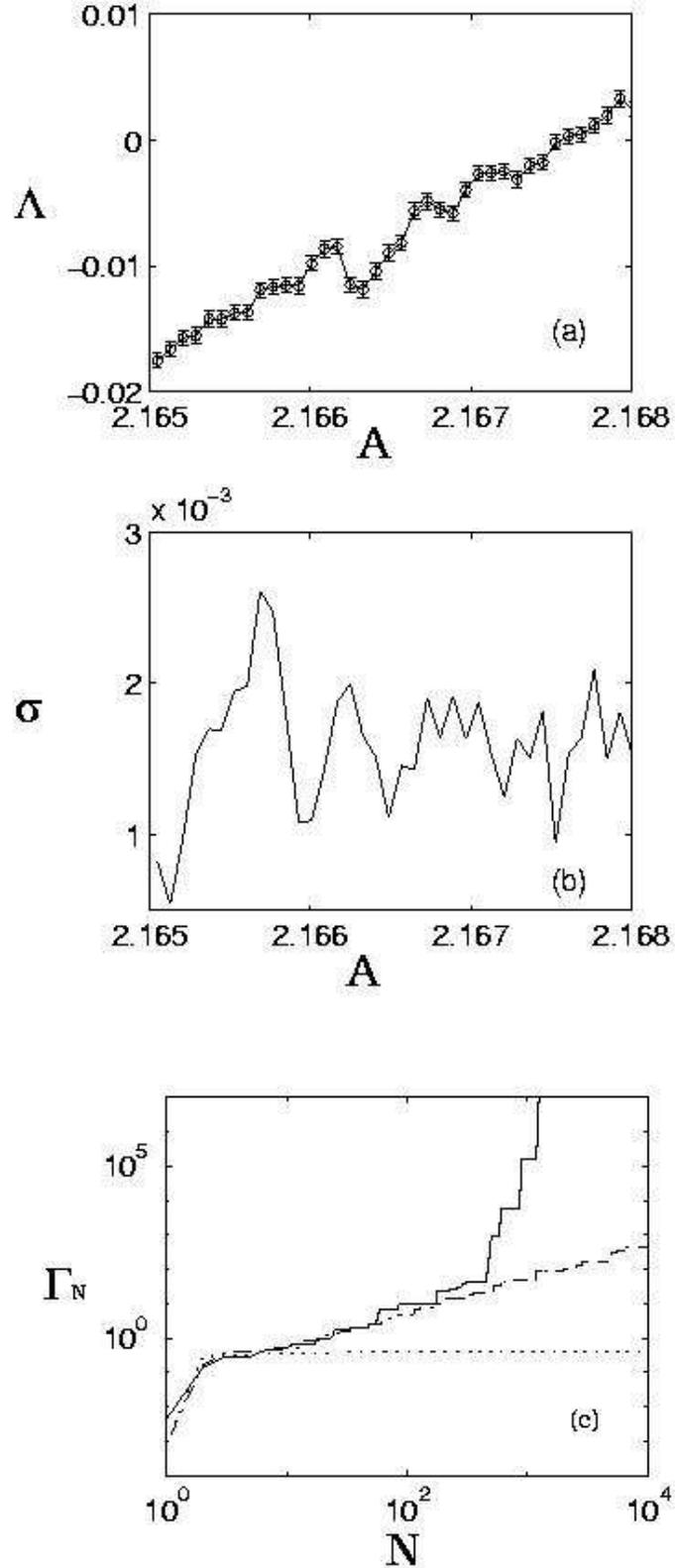, width=0.5\linewidth}}
\caption{Transition from doubled torus to SNA through gradual fractalization
mechanism in the region GF3: (a) the behaviour of the
Lyapunov exponent ($\Lambda$); (b) the variance ($\sigma$); (c) plot of phase
sensitivity function $\Gamma_N$ vs $N$ (dotted line corresponds to torus for $A$=1.85,
dashed line belongs to SNA for $A$=2.167 and solid line represents chaos for $A$=2.17).}
\end{figure}
%\newpage
\begin{figure}
\centerline{\epsfig {figure=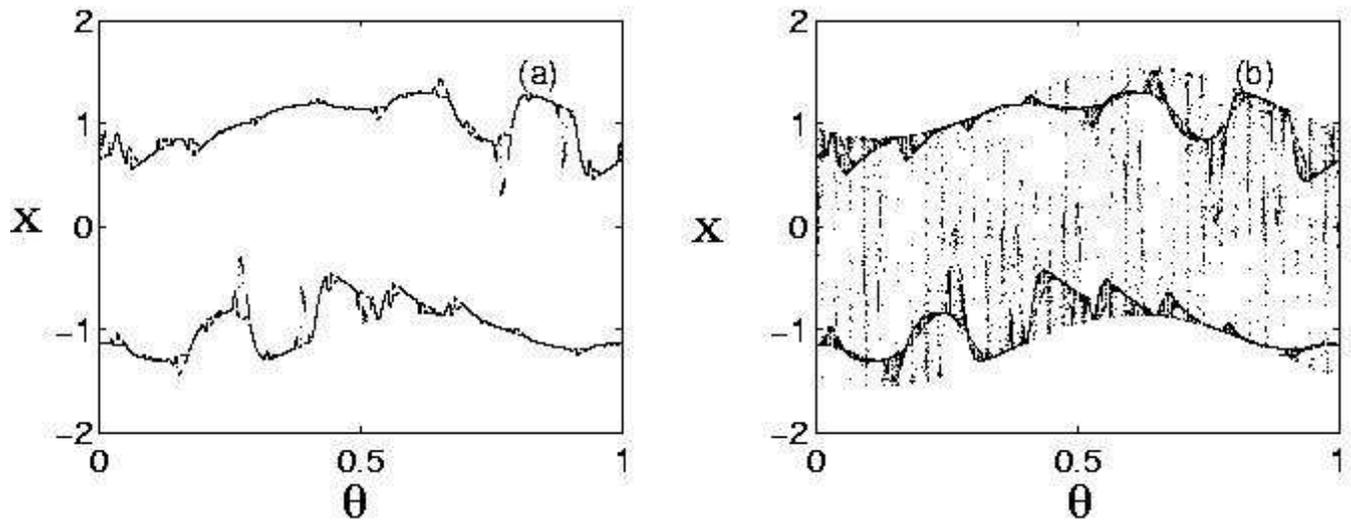, width=\linewidth}}
\caption{Projection of the  attractors of Eqs.~(3) for
$f$=0.35 in the $(x,\theta)$ plane  indicating
the transition from  quasiperiodic attractor  to chaotic attractor via SNA through
type--III intermittent mechanism:
(a) wrinkled attractor (period 2T) for $A$ = 2.135; (b)
SNA at $A$=2.14.}
\end{figure}
%\newpage
\begin{figure}
\centerline{\epsfig {figure=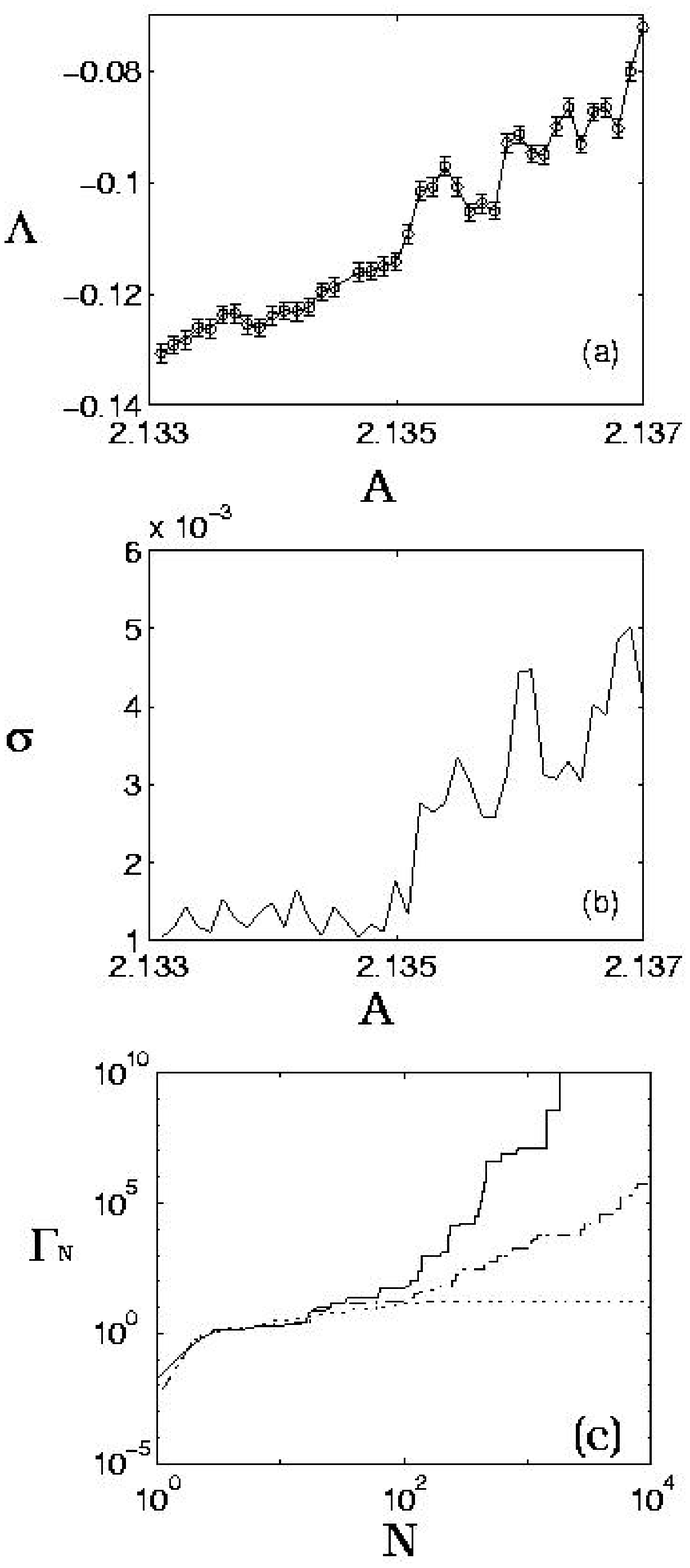, width=0.4\linewidth}}
\vskip 10pt
\caption{Transition from doubled torus to SNA through type--III intermittent
mechanism in the region S3: (a) the behaviour of the
Lyapunov exponent ($\Lambda$); (b) the variance ($\sigma$);(c) plot of phase
sensitivity function $\Gamma_N$ vs $N$ (dotted line corresponds to torus for $A$=1.83,
dashed line belongs to SNA for $A$= 2.14 and solid line represents  chaos for $A$=2.15).}
\end{figure}
%\newpage
\begin{figure}
\centerline{\epsfig {figure=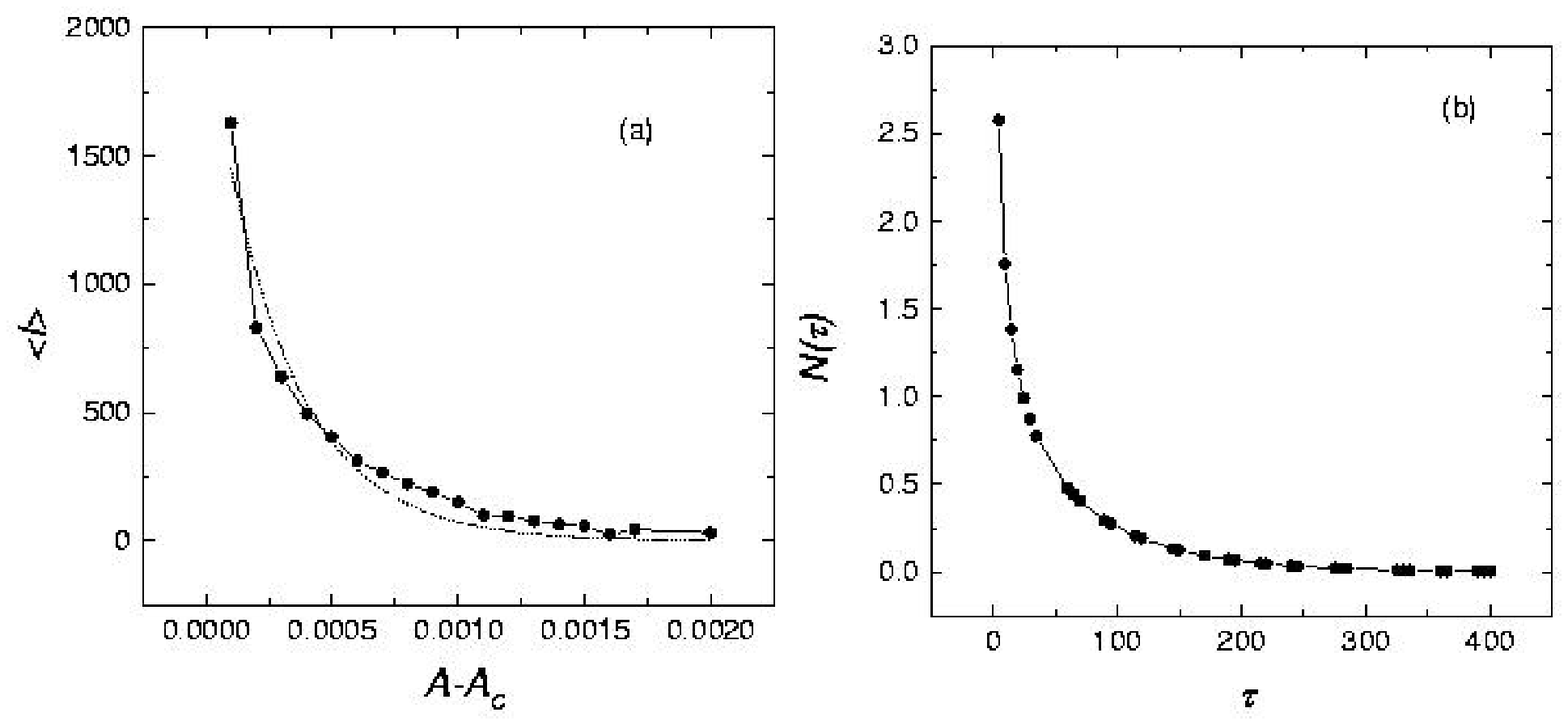, width=\linewidth}}
\caption { (a) Average  laminar length  ($<l>$) vs $(A-A_c)$ at $f$=0.35;
(b) Number of laminar periods $N(\tau)$ of duration $\tau$ 
in the case of transition through type-III intermittency.}
\end{figure}
%\newpage
\begin{figure}
\centerline{\epsfig {figure=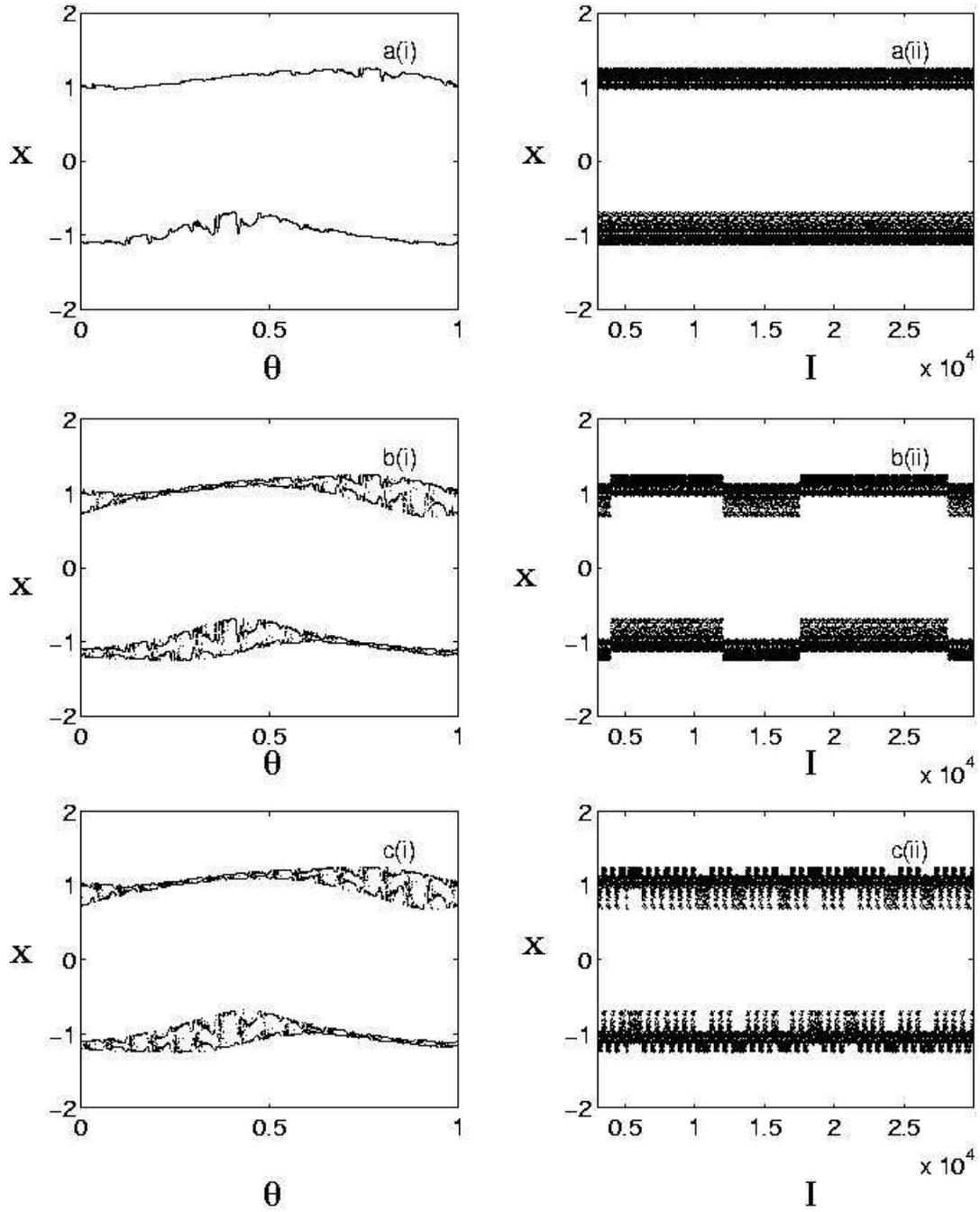, width=0.8\linewidth}}
\caption{Projection of the   attractors of Eqs.~(3) for
$f$=0.2 (i) in the $(x,\theta)$ plane; (ii) in the $(x,i)$ plane  indicating
the transition from  quasiperiodic attractor  to chaotic attractor via SNA through
crisis-induced intermittent mechanism:
(a) wrinkled attractor (period 2T) for $A$ = 2.138; (b)
SNA at $A$=2.1387; (c) SNA at $A$=2.1388.}
\end{figure}
\begin{figure}
\centerline{\epsfig {figure=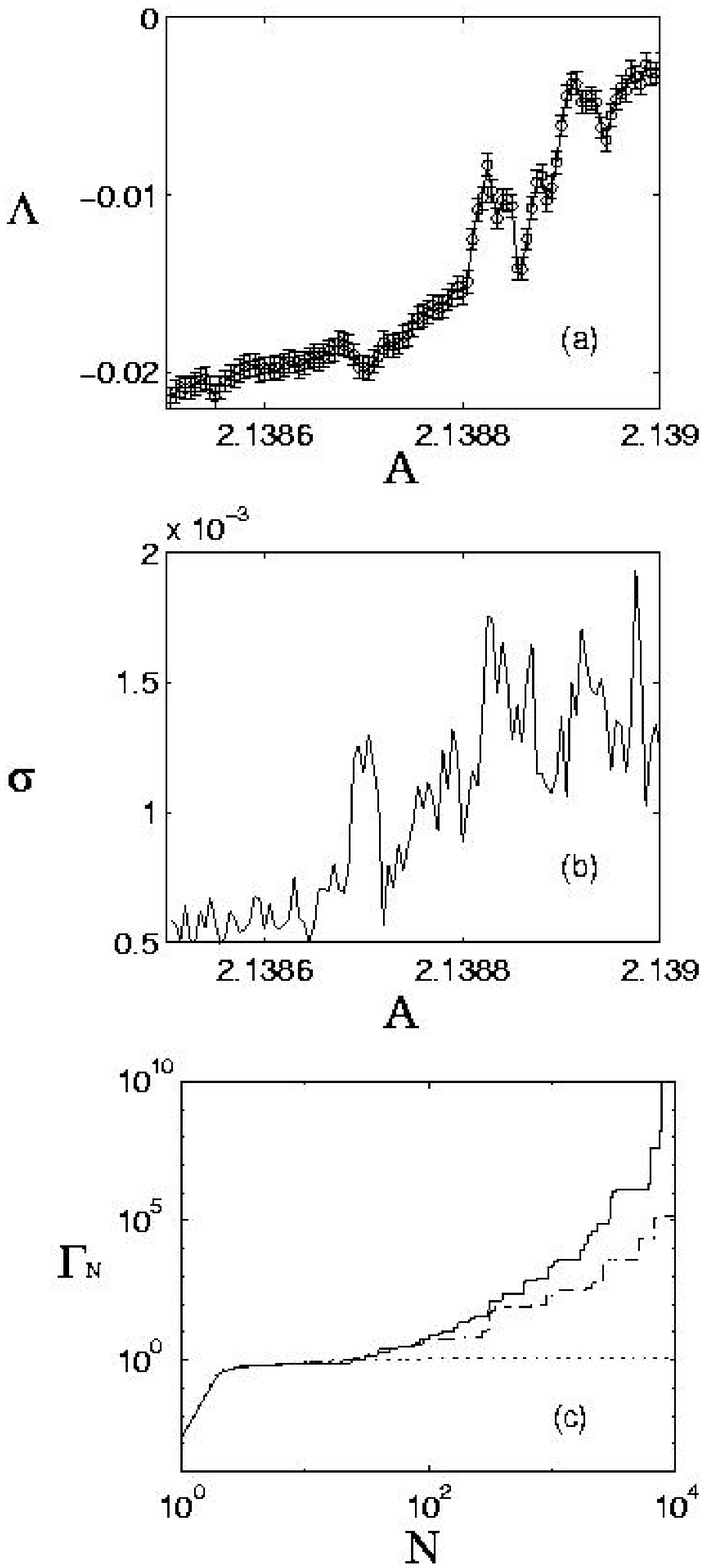, width=.4\linewidth}}
\caption{Transition from doubled torus to SNA through  crisis-induced 
intermittency
mechanism in the region IC: (a) the behaviour of the
Lyapunov exponent ($\Lambda$); (b) the variance ($\sigma$);(c) plot of phase
sensitivity function $\Gamma_N$ vs $N$ (dotted line corresponds to torus for $A$=1.83,
dashed line belongs to SNA for $A$=2.1388  and solid line represents  chaos for $A$=2.139).}
\end{figure}
%\newpage
\begin{figure}
\centerline{\epsfig {figure=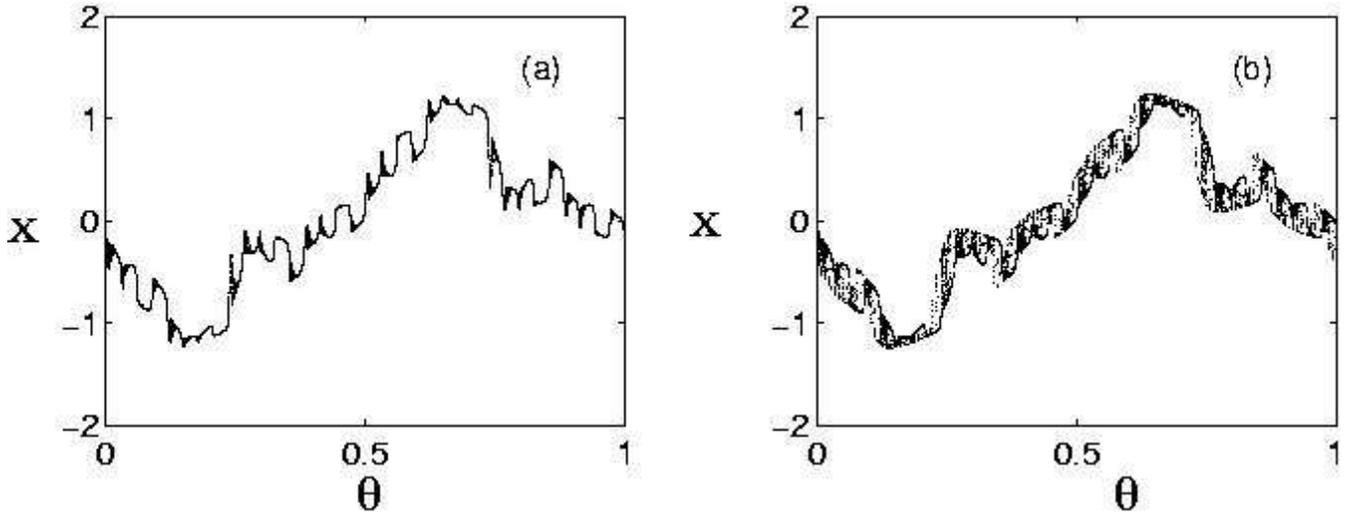, width=\linewidth}}
\caption{Projection of the attractors of Eqs.~(3) for
$f$=0.7 in the $(x,\theta)$ plane  indicating
the transition from  one-frequency quasiperiodic attractor to chaotic attractor via SNA through
gradual fractalization mechanism:
(a) wrinkled attractor (Period 1T) for $A$ = 1.25; (b)
SNA at $A$=1.265.}
\end{figure}
%\newpage
\begin{figure}
\centerline{\epsfig {figure=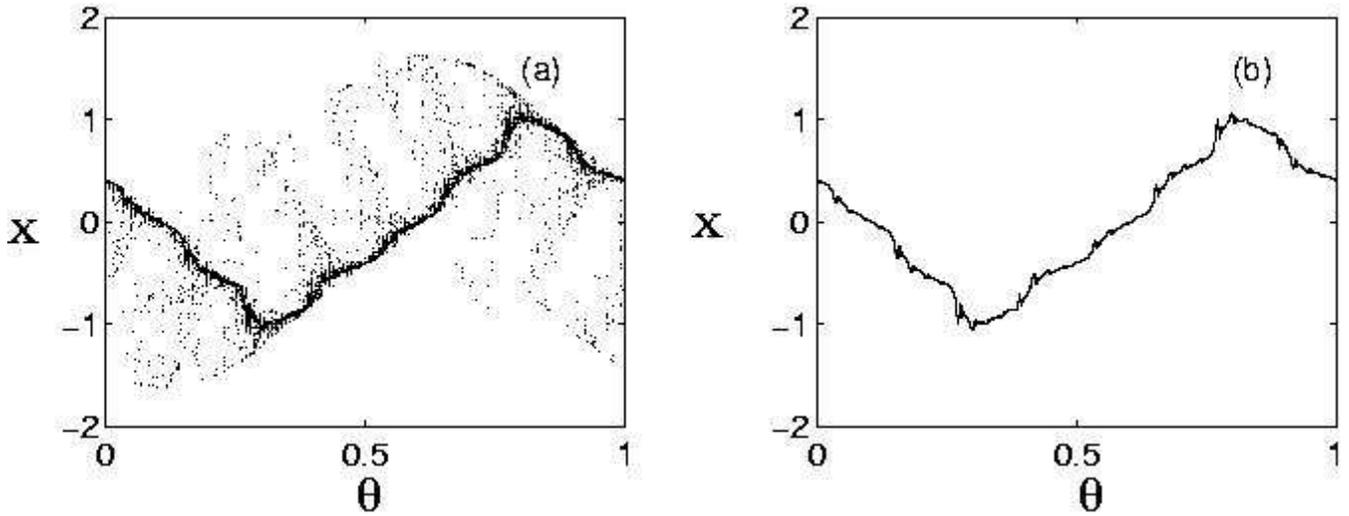, width=\linewidth}}
\caption{Projection of the  attractors of Eqs.~(3) for
$f$=0.7 in the $(x,\theta)$ plane  indicating
the transition from  one-frequency quasiperiodic attractor to chaotic attractor via SNA through
Type-I intermittency mechanism:
(a) intermittent SNA  for $A$ = 1.801685; (b)
torus  at $A$=1.8017.}
\end{figure}
%\newpage
\begin{figure}
\centerline{\epsfig {figure=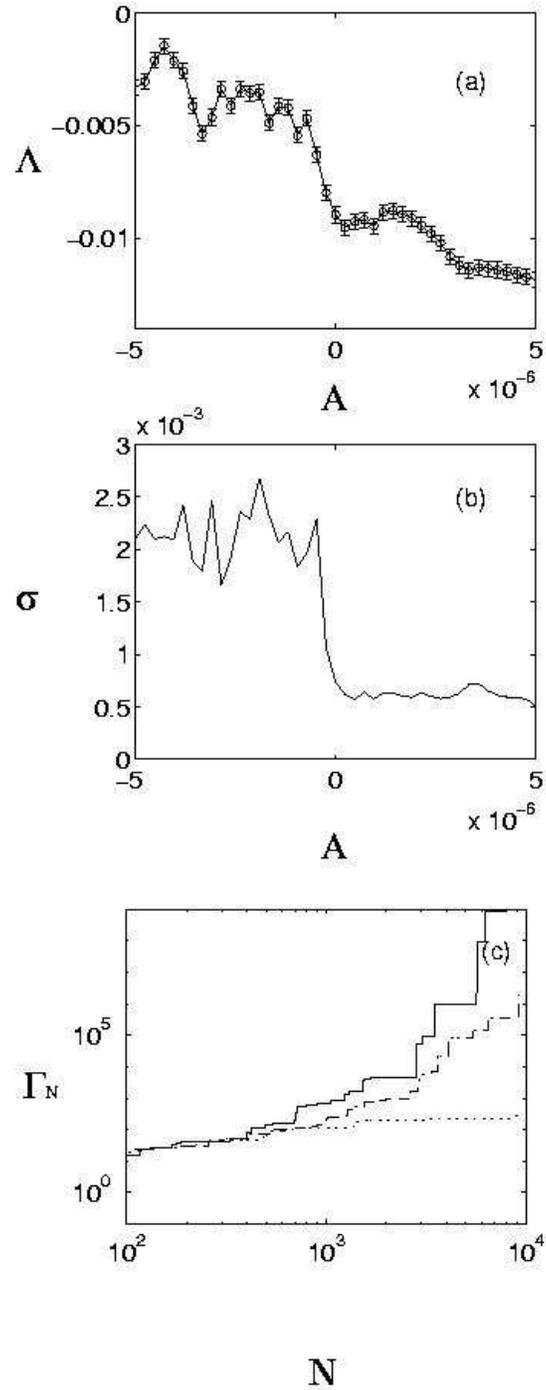, width=.4\linewidth}}
\caption{Transition from intermittent SNA to torus  through type I intermittent
mechanism in the region S1: (a) the behaviour of the
Lyapunov exponent ($\Lambda$); (b) the variance ($\sigma$);
(c) plot of phase
sensitivity function $\Gamma_N$ vs $N$ (dotted line corresponds to torus for $A$=1.8017,
dashed line belongs to SNA for $A$=1.801685 and solid line represents to chaos for $A$=1.8015).}
\end{figure}
\begin{figure}
\centerline{\epsfig {figure=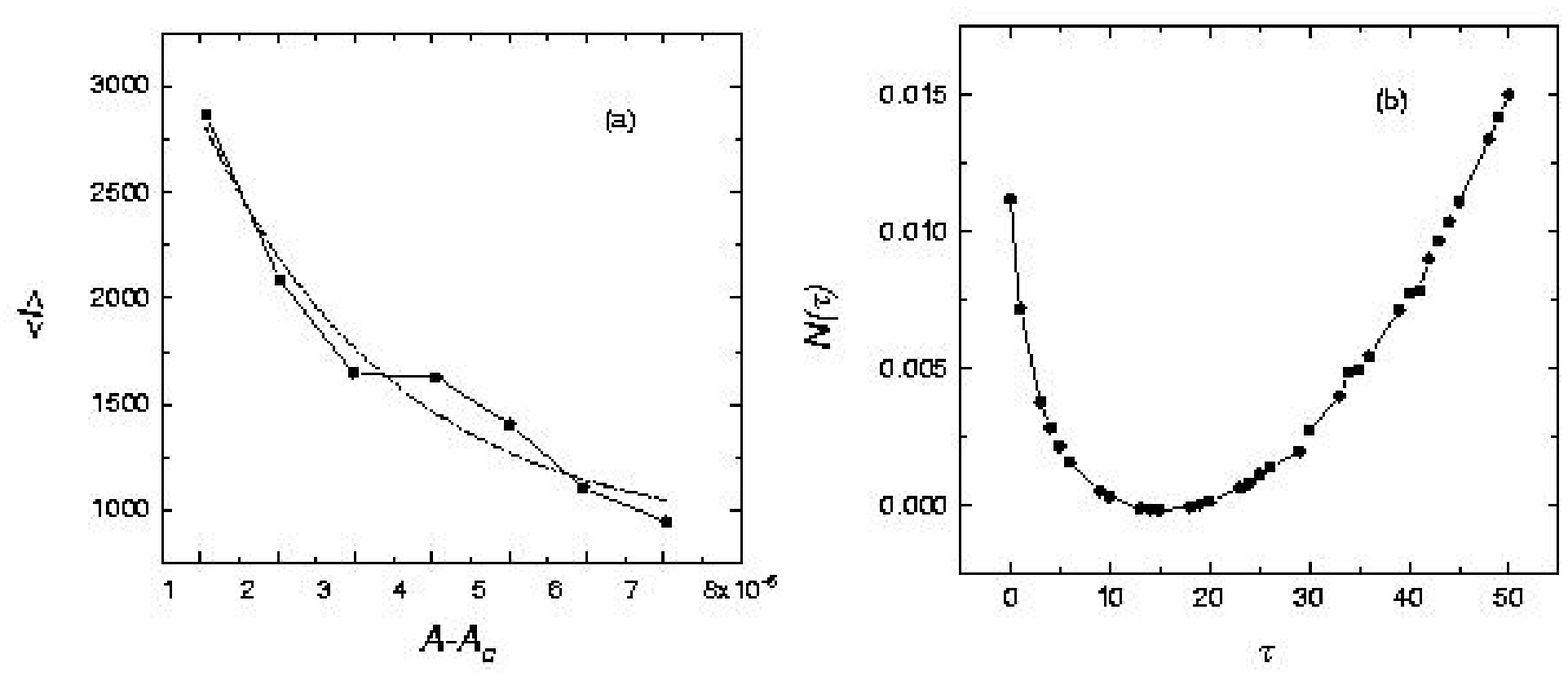, width=\linewidth}}
\caption {(a) Average  laminar length between ($<l>$) vs $(A-A_c)$ at $f$=0.7;
(b) Number of laminar periods $N(\tau)$ of duration $\tau$ 
in the case of transition through type-I intermittency.}
\end{figure}
\begin{figure}
\centerline{\epsfig {figure=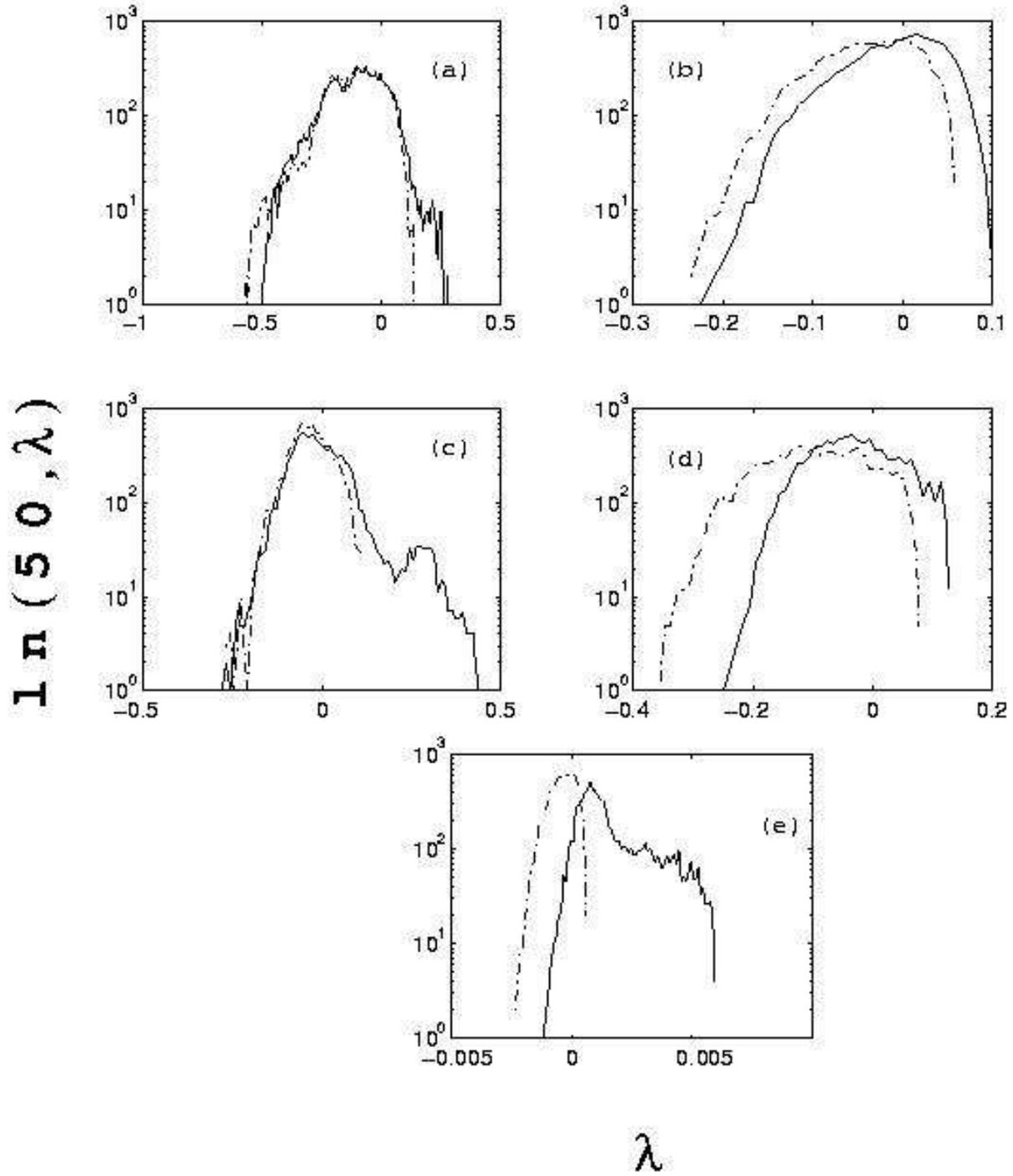, width=\linewidth}}
\caption{Distribution of finite--time Lyapunov exponents on 
SNAs created through (a) Heagy-Hammel, (b) gradual fractalization, (c) type III intermittency
and (d) crisis-induced intermittency  (e) type I intermittency. 
Solid and dashed lines correspond to SNA and torus distributions.}
\end{figure}
\newpage

\begin{table}
Table I Routes and mechanisms of the onset of various SNAs in the 
quasiperiodically forced cubic map
\begin{tabular}{l|l|l|l|l|l|l}
{Type of Route} & {Mechansim} & \multicolumn{4}{c|}{Characteristic properties} & { Figures } \\
\hline
\multicolumn{2}{c|}{} & {Lyapunov}&{Variance $\sigma$} &{The fraction of } & {Scaling law} &  \\
\multicolumn{2}{l|}{} & {exponent $\lambda$} & & {positive valued} & {$<l> $} &\\
\multicolumn{2}{l|}{}&   &     &{finite-time }& {$\sim $} & \\
\multicolumn{2}{l|}{}&     &     & {Lyapunov}&{$({A_c-A})^{\alpha}$}  &  \\
\multicolumn{2}{l|}{}&     &     & {exponent} & & \\
\multicolumn{2}{l|}{}&     &     & {($F_+{(N,\lambda)}$)} & & \\
\hline 
\multicolumn{7}{l}{ A. Interruption of torus doubling } \\
\hline
{1. Heagy-}& {Collision between a period- }&{Irregular in SNA} &{Small in torus}&{Decays} & & { Figs.~3}\\
{Hammel [8]}&{doubled torus and its}&{region \& smooth } & {\& large in SNA} & {exponentially} & &{\& 4} \\
&{unstable parent}& {in torus} & & & & \\
\hline
{2. Gradual }& {Torus gets increasingly } &{Increases slowly } &{No significant}&{Decays}& &{ Figs.~5}\\
{fractalization [9]}&{wrinkled and transforms} &{during the transi} &{changes}&{exponentially} & &{\& 6} \\
&{into a SNA without any} &{-tion from torus} & & & & \\
&{interaction with a nearby} & {to SNA} & & & & \\
&{unstable  periodic orbit} & & & & \\
\hline
{3. Type--III} & {During the transition from} & {Abrupt change } & {Abrupt increase} &{Power-law} & {$\alpha \sim 1.1$}&{Figs.~7,8} \\
{intermittency} &{torus doubled attractor to} & {during the} & {at the transi} &{variation} &{(see also} & {\& 9}\\
{[15]}&{SNA, a growth of subhar} &{transition from} &{-tion point} & &{eq.(5))} \\
&{-monic amplitude begins } & {torus to SNA} & & & \\
&{together with a decrease  } & & & &  &\\
&{in the size of the } & & & & \\
&{fundamental amplitude} & & & & \\ 
\hline
{4. Crisis-induced}&{Doubling of destroyed torus} &{Does not follow} & {Irregular variation} & {Decays} &  & {Figs.~10,} \\
{intermittency} & {involves a kind of sudden }&{uniform pattern} &{in SNA region} & {exponentially} & &{\& 11}  \\
 &{widening of the attractor} &  & & & \\
\hline
\multicolumn{7}{l}{B. Transition from one-frequency torus} \\
\hline
{1. Gradual} & {Torus gets increasingly} & {Increases slowly } & {No significant} & {Decays} &  &{Fig.~12} \\
{fractalization} & {wrinkled and transforms} & {during the transition } & {change} & {exponentially} & &  \\
{[9]} & {into a SNA} & {from torus to SNA} & & & & \\
\hline
{4. Type--I} & {Torus is eventually } & {Abrupt change } & {Abrupt increase} &{Power-law} & {$\alpha \sim
0.55$}&{Figs.~13,} \\
{intermittency} &{replaced by SNA } & {during the} & {at the transi} &{variation} &{(see also} &{14,\& 15} \\
{[14]}&{through an analog  } &{transition from} &{-tion point} & &{eq.(7))} &\\
&{of the saddle-node } & {torus to SNA} & & & \\
&{bifurcation } & & & & & \\
\end{tabular}
\end{table}
\end{document}